\hsize=14 cm \vsize=20.8 cm \tolerance=400
\hoffset=2cm\def\vec#1{{\bf #1}}
\font\trm = cmr10 scaled \magstep3
\font\srm = cmr10 scaled \magstep2

\voffset=2cm
\scriptscriptfont0 =\scriptfont0
\scriptscriptfont1 =\scriptfont1

\def\d{\partial}
\def\dh{\mathop{\vphantom{\odot}\hbox{$\partial$}}}
\def\dl{\dh^\leftrightarrow}
\def\sqr#1#2{{\vcenter{\vbox{\hrule height.#2pt\hbox{\vrule width.#2pt 
height#1pt \kern#1pt \vrule width.#2pt}\hrule height.#2pt}}}}
\def\w{\mathchoice\sqr45\sqr45\sqr{2.1}3\sqr{1.5}3\,}

\def\psq{{\overline{\psi}}}

\def\=d{\,{\buildrel\rm def\over =}\,}

\def\i3p{\p32\int d^3p}

\def\As{A\hbox to 1pt{\hss /}}
\def\np4{\int d^4p_1\cdots d^4p_{n-1}\, }

\def\supp{{\rm supp}\, }

\def\nx4{\int d^4x_1\ldots d^4x_n\, }

\def\kon#1#2{\vbox{\halign{##&&##\cr
\lower4pt\hbox{$\scriptscriptstyle\vert$}\hrulefill &
\hrulefill\lower4pt\hbox{$\scriptscriptstyle\vert$}\cr $#1$&
$#2$\cr}}}

\def\konv#1#2#3{\hbox{\vrule height12pt depth-1pt}
\vbox{\hrule height12pt width#1cm depth-11.6pt}
\hbox{\vrule height6.5pt depth-0.5pt}
\vbox{\hrule height11pt width#2cm depth-10.6pt\kern5pt
      \hrule height6.5pt width#2cm depth-6.1pt}
\hbox{\vrule height12pt depth-1pt}
\vbox{\hrule height6.5pt width#3cm depth-6.1pt}
\hbox{\vrule height6.5pt depth-0.5pt}}
\def\konu#1#2#3{\hbox{\vrule height12pt depth-1pt}
\vbox{\hrule height1pt width#1cm depth-0.6pt}
\hbox{\vrule height12pt depth-6.5pt}
\vbox{\hrule height6pt width#2cm depth-5.6pt\kern5pt
      \hrule height1pt width#2cm depth-0.6pt}
\hbox{\vrule height12pt depth-6.5pt}
\vbox{\hrule height1pt width#3cm depth-0.6pt}
\hbox{\vrule height12pt depth-1pt}}

\def\konw#1#2#3{\hbox{\vrule height12pt depth-1pt}
\vbox{\hrule height12pt width#1cm depth-11.6pt}
\hbox{\vrule height6.5pt depth-0.5pt}
\vbox{\hrule height12pt width#2cm depth-11.6pt \kern5pt
      \hrule height6.5pt width#2cm depth-6.1pt}
\hbox{\vrule height6.5pt depth-0.5pt}
\vbox{\hrule height12pt width#3cm depth-11.6pt}
\hbox{\vrule height12pt depth-1pt}}

\def\i{{\rm int}}

\def\m3{{\mu_1\mu_2\mu_3}}

\def\p{{(+)}}

\nopagenumbers
\vbox to 1.5cm{ }
\centerline{Preprint: hep-th/9807078}
\centerline{DESY 98-090}
\vbox to 1.5cm{ }
\centerline{\trm A local (perturbative) construction of observables}
\vskip 0.5cm
\centerline{\trm in gauge theories: the example of QED}
\vskip 2cm
\centerline{\srm M. D\"utsch (*) and K. Fredenhagen}\vskip 0.5cm
\centerline{\it II. Institut f\"ur Theoretische Physik}
\centerline{\it Universit\"at Hamburg}
\centerline{\it Luruper Chaussee 149}
\centerline{\it D-22761 Hamburg, Germany}\vskip 3cm
{\bf Abstract.} - Interacting fields can be constructed as formal power series
in the framework of causal perturbation theory. The local field algebra
$\tilde {\cal F}({\cal O})$ is obtained without performing the adiabatic limit;
the (usually bad) infrared behavior plays no role. To construct the observables
in gauge theories we use the Kugo-Ojima formalism; we define the 
BRST-transformation $\tilde s$ as a graded derivation 
on the algebra of interacting
fields and use the implementation of $\tilde s$ by the Kugo-Ojima operator
$Q_{\rm int}$. Since our treatment is local, the operator $Q_{\rm int}$
differs from the corresponding operator $Q$ of the free theory. We prove
that the Hilbert space structure present in the free case is stable under
perturbations. All assumptions are shown to be satisfied in QED.
\vskip 0.5cm
{\bf PACS.} 11.15.-q Gauge field theories, 11.15.Bt General properties of
perturbation theory 
\vskip 1.5cm
\centerline{(*) Work supported by the Alexander von Humboldt Foundation}
\centerline{e-mail: duetsch@puls01.desy.de, fredenha@x4u2.desy.de}
\vfill\eject
\pageno=1
\headline={\tenrm\ifodd\pageno\hss\folio\else\folio\hss\fi}
\footline={\hss}
{\trm 1. Introduction}
\vskip 1cm
The quantization of gauge theories is a longstanding problem of theoretical 
physics. Since the works of Tomonaga, Schwinger, Feynman and
Dyson in the late fourties the problem is solved for QED from a pragmatic point
of view: the predictions (e.g. on the magnetic moment of the electron) are 
in perfect agreement with experiment. In the sixties and seventies the 
quantization of nonabelian gauge theories was developed by Feynman [F], 
Faddeev-Popov [FP], t'Hooft, Becchi-Rouet-Stora [BRS], Kugo-Ojima [KO] and
others. Weinberg and Salam proposed to base the theory of electroweak 
interactions on a spontaneously broken gauge model, which has survived
the last thirty years. 

The ultraviolet divergences appearing in quantum field theory can be removed by
various renormalization methods. 
An elegant method is causal perturbation theory
which was developed by Epstein and Glaser [EG] on the basis of ideas due to
St\"uckelberg and Bogoliubov and Shirkov [BS]. 
However, the {\it infrared problem}
is only partially solved. One aspect is that charged 
particles cannot be eigenstates 
of the mass operator (they have to be "infraparticles" [Sch, Bu2]). 
Another aspect
of the infrared problem are the divergences which appear in the adiabatic limit
$g\rightarrow$ const. of the S-matrix, where $g$ is a space-time dependent
coupling 'constant'. In QED these divergences are logarithmic and cancel in 
the cross section. (This is proven at least at low orders of the perturbation 
series [S].) Moreover, Blanchard and 
Seneor [BlSe] proved that the adiabatic limit
of Green's and Wightman functions exists for QED.
But in nonabelian gauge theories the divergences are worse. Perturbation 
theory seems to be unable to describe the long distance properties of these 
models ("confinement"). There is the hope that these two aspects of the
infrared problem are directly connected. (See e.g. Scharf in
[S], sect. 3.12, where the existence 
and uniqueness of the adiabatic 
limit of the S-matrix is proposed  to be a good criterion 
for a selection of physical states.)

The main message of the present paper is that a {\it local} construction of 
the observables in gauge theories is possible without performing the adiabatic 
limit. Hence the infrared divergences do not occur in the 
construction of the model.
They rather appear on the level of the long distance properties of the theory.
We hope that due to its local character, our construction can be 
generalized to curved space-times, continuing the program of [BFK,BF].

The quantization of gauge fields in a renormalizable gauge
requires an indefinite metric space. Afterwards, one has to prove that
the Wightman distributions of gauge invariant fields fulfil the condition of
positivity. In [DHKS1,S] the free Kugo-Ojima charge
$Q$ which implements the BRST-transformation of free fields is used to
select the physical Hilbert space. 
But the commutator of $Q$ with the interacting
gauge invariant fields vanishes only up to a divergence. One may expect that in
the adiabatic limit the positivity is satified, but a proof in the nonabelian
case seems to be rather hard. Here we adopt another point of view which avoids
the discusion of the adiabatic limit. Our way out is to work with the 
interacting Kugo-Ojima charge $Q_{\rm int}$ [KO],
which implements the BRST-transformation [BRS] of the interacting fields. 
By means of the Ward identities we 
prove that the commutator of $Q_{\rm int}$ with
the gauge invariant interacting fields of QED is in fact zero. The infrared
divergences which remained an open 
problem in the Kugo-Ojima formalism are absent
in our treatment. However, since $Q_{\rm int}\not= Q$ 
before the adiabatic limit,
the construction of the physical Hilbert space cannot be done on the level of
free fields. We show that the physical Hilbert space of the interacting model
is obtained as a deformation of the physical Hilbert space of 
the free theory. Here we adopt ideas from deformation 
quantization as developed by Bordemann and Waldmann [BW].

The paper is organized as follows. In the next section we study 
the interacting fields in the framework of causal perturbation theory [BS,EG].
They are formal power series of (unbounded)
operators in the Fock space of free fields. We point out that up to unitary
equivalence the interacting fields depend on the interaction Lagrangian 
only {\it locally}. In section 3 we specialize to QED and 
compute commutators of interacting fields to all orders. 
Thereby we essentially use
the Ward identities, which are proven in appendix B.
 
In section 4 we turn to the problems specific to gauge theories, 
the elimination of the
unphysical fields and the mentioned positivity. We give a general {\it local}
construction of the {\it observables} in gauge 
theories and of the {\it physical 
Hilbert space} in which the observables 
are faithfully represented. We prove that 
this structure is stable if we replace the free fields by the interacting ones.

This general construction relies on some assumptions. 
They are verified in the case
of QED in section 5. The main problem is the construction of the 
interacting Kugo-Ojima charge $Q_{\rm int}$. 
To avoid a volume divergence we embed
the algebra of interacting fields on an arbitrary finitely extended region
into the corresponding algebra over a spacetime with compact spatial sections.
This does not change the algebraic relations. We expect that also the
corresponding Hilbert space representations of the local algebras of
observables are unitarily equivalent, but this remains to be proven.
The technical details of the free quantum gauge fields on 
the spatially compactified Minkowski space are written down in appendix A.
\vskip 1cm
{\trm 2. Perturbative Construction of Interacting Fields}
\vskip 1cm
In the framework of causal perturbation theory 
[BS,EG,St3,S,BF], interacting fields
can be constructed as formal power series of operator valued distributions on a
dense invariant domain ${\cal D}$ in the Fock space of incoming fields. The
interacting fields $A_{{\rm int}\>{\cal L}}(x)$ ($A$ is a Wick polynomial
of incoming fields)
depend on an interaction Lagrangian ${\cal L}$
which is a Wick polynomial of incoming fields 
with test functions $g\in {\cal D}
({\bf R}^4)$ as coefficients so that the interaction 
is switched on only within a
finitely extended region of spacetime.

The crucial observation is that the dependence of the interacting fields on the
interaction Lagrangian is {\it local}, in the sense that
given a causally complete finitely extended open
spacetime region ${\cal O}$, Lagrangians ${\cal L}_1$ and
${\cal L}_2$ which differ only within a closed region which does not
intersect the closure of ${\cal O}$, lead to unitarily equivalent fields
within ${\cal O}$,
i.e. there exists a unitary formal power series $V$ of operators on ${\cal D}$
such that
$$VA_{{\rm int}\>{\cal L}_1}(x)V^{-1}=
A_{{\rm int}\>{\cal L}_2}(x),\quad\quad\quad
\forall x\in {\cal O},\eqno(2.1)$$ 
and $V$ does not depend on $A$ [BF]. This 
property (2.1) is a direct consequence of 
causality, which we are now going to explain.

Building blocks for the construction of interacting fields are the time ordered
products $T(A_1(x_1)...A_n(x_n))$ of Wick polynomials of free fields. They are 
multilinear (with ${\cal C}^\infty$ functions 
as coefficients) symmetrical operator
valued distributions on the dense domain ${\cal D}$ which satisfy the causal 
factorization property
$${\bf (Causality)}\quad T(A_1(x_1)...A_n(x_n))=
T(A_1(x_1)...A_k(x_k))T(A_{k+1}(x_{k+1})...A_n(x_n))\eqno(2.2)$$
if $x_j\not\in \bar 
V_{+}+x_i,\>i=1,...,k,\>j=k+1,...,n$, where $\bar V_{+}$ 
is the closed forward light cone in Minkowski space.

Causality (2.2) and symmetry determine the time ordered products on the set of 
pairwise different points. Moreover, if the time ordered products of less than
$n$ factors are everywhere defined, the time ordered product of $n$ factors is
uniquely determined up to the total 
diagonal $x_1=...=x_n$. Thus renormalization 
amounts to an extension, for every $n$, of time ordered products to the total
diagonal.
This extension is always possible, and it can be 
done such that the conditions of 
Poincare covariance (w.r.t. some unitary positive energy representation $U$ of 
the Poincare group ${\cal P}_+^\uparrow$)
$${\bf (N1)}\quad {\rm Ad}\,U(L)(T(A_1(x_1)...A_n(x_n)))=
T({\rm Ad}\,U(L)(A_1(x_1))...{\rm Ad}\,U(L)(A_n(x_n))),\quad L\in 
{\cal P}_+^\uparrow\eqno(2.3)$$
and of unitarity hold,\footnote{$^1$}{We work throughout with the conventions
of [EG], not with the ones of [S].}
$${\bf (N2)}\quad\quad\quad T(A_1(x_1)...A_n(x_n))^+=\sum_{P\in {\rm Part}\,
\{1,...,n\}}(-1)^{|P|+n}\prod_{p\in P}T(A_i(x_i)^+,\,i\in p).\eqno(2.4)$$
($^+$ means the adjoint on ${\cal D}$, 
$\>(\phi,B(f)\psi)=(B^+(\bar f)\phi,\psi),
\quad\phi,\psi\in {\cal D}$.) The generating functional 
for the time ordered products
is the $S$-matrix $S({\cal L}),\>{\cal L}=\sum_{i=1}^N g_iA_i,\,g_i\in
{\cal D}({\bf R}^4)$
$$S({\cal L})=\sum_{n=0}^\infty {i^n\over n!}\int d^4x_1...d^4x_n\,
T({\cal L}(x_1)...{\cal L}(x_n)),\eqno(2.5)$$
i.e.
$$T(A_{i_1}(x_1)...A_{i_n}(x_n))={\delta^n\over i^n\delta g_{i_1}(x_1)...\delta
g_{i_n}(x_n)}S({\cal L})\vert_{g_1=...=g_N=0}.\eqno(2.6)$$
Finally, the interacting field $A_{{\rm int}\>{\cal L}}$ 
corresponding to the Wick
polynomial $A$ of the free fields, is defined by [BS,EG,DKS1]
$$A_{{\rm int}\>{\cal L}}(x)={\delta\over i\delta h(x)}
S({\cal L})^{-1}S({\cal L}+hA)
\vert_{h=0}.\eqno(2.7)$$
By inserting (2.5) into (2.7) one obtains the perturbative expansion of the
interacting fields
$$A_{{\rm int}\>{\cal L}}(x)=A(x)+\sum_{n=1}^{\infty}{i^n\over n!}
\int d^4x_1...d^4x_n\,R({\cal L}(x_1)...{\cal L}(x_n);A(x)),\eqno(2.8)$$
with the 'totally retarded products'
$$R(A_1(x_1)...A_n(x_n);A(x))\=d \sum_{I\subset \{1,...,n\}}(-1)^{|I|}
\bar T(A_i(x_i),\,i\in I)T(A_j(x_j),\,j\in I^c,A(x)),\eqno(2.9)$$
where $I^c\=d \{1,...,n\}\setminus I$ and $\bar T$ 
denotes the 'antichronological
product'. The corresponding generating functional is the inverse $S$-matrix
$$S({\cal L})^{-1}=\sum_{n=0}^\infty {(-i)^n\over n!}\int d^4x_1...d^4x_n\,
\bar T({\cal L}(x_1)...{\cal L}(x_n)),$$
and the antichronological products can be obtained 
from the time ordered products
by the usual inversion of a formal power series, namely the r.h.s. of (2.4).

The arbitrariness in the extensions of time ordered products to coinciding
points can be further restricted. Let $\phi_1,...,\phi_N$ be the free fields in
terms of which the model is defined, which satisfy the linear field equation
$$\sum_jD_{ij}\phi_j=0,\eqno(2.10)$$
(where $D_{ij}$ is a matrix whose entries are differential operators such that
$D$ is a relativistic invariant hyperbolic differential operator with a
unique solution of the Cauchy problem) and with C-number commutators
$$[\phi_j(x),\phi_k(y)]=i\Delta_{jk}(x-y),\quad\quad \Delta_{jk}=
\Delta_{jk}^{\rm ret}-\Delta_{jk}^{\rm av}\eqno(2.11)$$
($\Delta_{jk}^{\rm ret,\,av}$ retarded resp. advanced 
Green's function of $D_{ij}$, i.e. ${\rm supp}\>\Delta_{jk}^{\rm ret,\,av}
\subset\bar V_{+,-}$).
We define to every Wick polynomial $A$ the sub Wick polynomials ${\d A\over
\d\phi_j}$ by \footnote{$^2$}{If $A$ contains 
{\it derivated} free fields the definition
(2.12) is replaced by
$$[A(x),\phi_k(y)]=i\int d^4z\,
\sum_j{\delta A(x)\over\delta\phi_j(z)}\Delta_{jk}(z-y)$$
and similar modifications appear in the following formulas.}
$$[A(x),\phi_k(y)]=i\sum_j{\d A\over\d\phi_j}(x)\Delta_{jk}(x-y).\eqno(2.12)$$
We then require
$${\bf (N3)}\quad [T(A_1(x_1)...A_n(x_n)),
\phi_j(x)]=i\sum_{k=1}^n\sum_lT(A_1(x_1)...
{\d A_k\over\d\phi_l}(x_k)...A_n(x_n))\Delta_{lj}(x_k-x)\eqno(2.13)$$
and (cf. [St2])
$${\bf (N4)}\quad \sum_jD_{ij}^x
T(A_1(x_1)...A_n(x_n)\phi_j(x))=i\sum_{k=1}^nT(A_1(x_1)...
{\d A_k\over\d\phi_i}(x_k)...A_n(x_n))\delta (x_k-x).\eqno(2.14)$$
The first condition means that the time ordered product is determined up to a
C-number by the time ordered products of sub Wick polynomials, whereas the 
second condition defines uniquely the time ordered products with additional 
free field factors once it is given away from the diagonal. 
It translates into the differential equation
$${\bf (N4')}\quad \sum_jD_{ij}^xR(A_1(x_1)...A_n(x_n);\phi_j(x))=i\sum_{k=1}^n
R(A_1(x_1)...\hat k...A_n(x_n);
{\d A_k\over\d\phi_i}(x_k))\delta (x_k-x)\eqno(2.15)$$
for the totally retarded products (2.9), where the hat means the omission of
the corresponding factor. By means of (2.8) we see that the
requirement (2.14) implies the field equation for the interacting field
$\phi_{{\rm int}\>j,{\cal L}}$,
$$\sum_jD_{ij}\phi_{{\rm int}\>j,{\cal L}}(x)=
-\Bigl( {\d {\cal L}\over\d\phi_i}\Bigr)
_{{\rm int}\>{\cal L}}(x).\eqno(2.16)$$

The remaining arbitrariness is the freedom in the extension of the expectation
values $\omega (T(A_1(x_1)...A_n(x_n)))$, where $\omega$ is some state
(e.g. the vacuum), to the diagonal. This freedom consists in adding a 
distribution with
support on the diagonal. Its form is restricted by covariance and by the
requirement that the degree of the singularity at the diagonal, measured 
in terms of Steinmann' scaling degree [Ste,BF], may not be increased by 
the extension.

The requirements (2.13) and (2.14) are purely 
algebraic normalization conditions 
for the time ordered products resp. the interacting fields. 
They are independent
of the choice of some state and, hence, are well suited for the generalization
to curved spacetimes.

For later purpose we are going to list some properties of the totally retarded
products (2.9). By means of causality (2.2) one easily finds that they have 
totally retarded support,
$${\rm supp}\>R(A_1(x_1)...A_n(x_n);A(x))
\subset\{(x_1,...x_n;x)\mid x_i\in x+\bar V_-,\,\forall i=1,...n\}.
\eqno(2.17)$$
This means that the interacting fields $A_{{\rm int}\>{\cal L}}(x)$ 
(2.7-8) depend only on the
interaction in the past of $x$, i.e. solely on ${\cal L}\vert_{x+\bar V_-}$.

The following lemma describes the structure of the totally 
retarded products with 
a free field factor.

{\bf Lemma 1}: Let $(\phi_j)_j$ be free fields and the normalizations fulfil
{\bf (N4)}. Then
$${\bf (A)}\quad R(A_1(x_1)...A_n(x_n);\phi_i(x))=i\sum_{k=1}^n\sum_l
\Delta_{il}^{\rm ret}(x-x_k)
R(A_1(x_1)...\hat k...A_n(x_n);{\d A_k\over\d\phi_l}(x_k)),$$
$${\bf (B)}\quad R(A_1(x_1)...A_{n-1}(x_{n-1})\phi_j(y);A(x))=
i\sum_{m=1}^{n-1}\sum_h
\Delta_{jh}^{\rm av}(y-x_m)R(A_1(x_1)...{\d A_m\over\d\phi_h}(x_m)...$$
$$...A_{n-1}(x_{n-1});A(x))
+i\sum_h\Delta_{jh}^{\rm av}(y-x)R(A_1(x_1)...A_{n-1}(x_{n-1});{\d A\over
\d\phi_h}(x))$$
Thereby note
$${\bf (C)}\quad R(A_1(x_1)...A_n(x_n);A(x))=0\quad 
{\rm if}\>n\geq 1\>{\rm and}\>{\rm some}\>A_j\>(1\leq j\leq n)\>
{\rm or}\>A\>{\rm is}\>{\rm a}\>{\rm C-number}.$$
 
{\it Proof}: 
The last statement {\bf (C)} is easily obtained from the definition (2.9)
and, if $A$ is the C-number, by taking $\sum_{I\subset \{1,...,n\}}
\bar T(A_i(x_i),\,i\in I)T(A_j(x_j),\,j\in I^c)=0$ into account.
Alternatively  one can argue by means of {\bf (N1)} and the translation 
invariance of C-number fields that the non-validity of 
{\bf (C)} would contradict
the support property (2.17).

{\bf (A)} Due to
$$D_{ij}\Delta_{jk}^{\rm ret,\,av}(x)=\delta_{ik}\delta(x),\eqno(2.18)$$
the expression {\bf (A)} is a solution of the hyperbolic 
differential equation (2.15).
Moreover, it is the only solution which fulfills the support property (2.17).

To prove {\bf (B)} we note that {\bf (N4)} implies
$$D_{ij}^yR(A_1(x_1)...A_{n-1}(x_{n-1})\phi_j(y);A(x))=i\sum_{m=1}^{n-1}
\delta (y-x_m)R(A_1(x_1)...{\d A_m\over\d\phi_i}(x_m)...$$
$$...A_{n-1}(x_{n-1});A(x))+
i\delta (y-x)R(A_1(x_1)...A_{n-1}(x_{n-1});
{\d A\over\d\phi_i}(x)),\eqno(2.19)$$
analogously to (2.15). Again there is only one 
solution of (2.19) which respects
(2.17), namely {\bf (B)}. $\quad\w$

In the next section we will compute commutators of 
interacting fields by means of

{\bf Proposition 2}: The (anti-) 
commutator of two interacting fields can be
written in the form
$$[A^1_{{\rm int}\>{\cal L}}(x),A^2_{{\rm int}\>{\cal L}}(y)]_\mp =
-\sum_{n=0}^{\infty}{i^n\over n!}\int d^4y_1...d^4y_n$$
$$\Bigl\{ R({\cal L}(y_1)...{\cal L}(y_n)A^1(x);
A^2(y))\mp R({\cal L}(y_1)...{\cal L}(y_n)A^2(y);A^1(x))\Bigr\}.\eqno(2.20)$$
The anticommutator appears only if $A^1$ and $A^2$ have either both an odd
number of spinor or both an odd number of ghost fields.\footnote{$^3$}{We work 
with the convention that the Fock space of the (free) incoming fields is 
the tensor product of the photon, spinor and ghost Fock spaces (see (A.15)).
Hence a free spinor fields {\it commutes} with a free ghost field.}

{\it Proof}: Due to $S({\cal L}+hA)^{-1}S({\cal L}+hA)=1$ 
we can write (2.7) in an
alternative way
$$A_{{\rm int}\>{\cal L}}(x)=-{\delta\over i\delta h(x)}
S({\cal L}+hA)^{-1}S({\cal L})
\vert_{h=0}.\eqno(2.21)$$
and, hence, we get
$$A^1_{{\rm int}\>{\cal L}}(x)A^2_{{\rm int}\>{\cal L}}(y)={\delta^2\over
\delta h_1(x)\delta h_2(y)}S({\cal L}+h_1A^1)^{-1}S({\cal L}+h_2A^2)
\vert_{h_1=0=h_2}.\eqno(2.22)$$
Next we note that the first term on the r.h.s. of (2.20) is equal to
$$-{\delta^2\over i^2\delta h_1(x)\delta h_2(y)}S({\cal L}+h_1A^1)^{-1}
S({\cal L}+h_1A^1+h_2A^2)\vert_{h_1=0=h_2}.\eqno(2.23)$$
Therefore, the assertion (2.20) is equivalent to
$$[A^1_{{\rm int}\>{\cal L}}(x),A^2_{{\rm int}\>{\cal L}}(y)]=
{\delta^2\over\delta h_1(x)\delta h_2(y)}[S({\cal L}+h_1A^1)^{-1}
S({\cal L}+h_1A^1+h_2A^2)-$$
$$-S({\cal L}+h_2A^2)^{-1}
S({\cal L}+h_1A^1+h_2A^2)]\vert_{h_1=0=h_2},\eqno(2.24)$$
which holds obviously true by (2.22). If $A^i$ is fermionic the corresponding
test function $h_i$ is Grassmann valued (i.e. an anticommuting
C-number, see e.g. [S],
appendix D) and, hence, the commutators possibly turn 
into anticommutators. $\quad\w$

By means of the support property (2.17) we immediately see from (2.20)
that the interacting fields are local
$$[A^1_{{\rm int}\>{\cal L}}(x),A^2_{{\rm int}\>{\cal L}}(y)]_\mp =0
\quad\quad {\rm if}\quad\quad (x-y)^2<0.\eqno(2.25)$$
Of course this can also be proven in a more direct way.
Note that Proposition 2 also provides a 
decomposition of the commutator into a retarded (i.e. $(x-y)\in\bar V_+$)
and advanced part ($(x-y)\in\bar V_-$).
\vfill
\eject
{\trm 3. Commutators of Interacting Fields in QED}
\vskip 1cm
In QED the interaction is given by
$${\cal L}(x)=g(x):\psq (x)\gamma_\mu A^\mu (x)\psi (x):,
\quad\quad g\in {\cal D}
({\bf R}^4),\eqno(3.1)$$
where $A^\mu$ is the free photon field and $\psq,\,\psi$ are 
the free spinor fields. In addition,
we introduce a pair $u,\,\tilde u$ of free, anticommuting ghost fields
$$\w u=0=\w\tilde u,\quad\>\{u(x),u(y)\}=0,\quad\>
\{\tilde u(x),\tilde u(y)\}=0,\quad\>\{u(x),\tilde u(y)\}=
-iD (x-y),\eqno(3.2)$$
where $D$ is the massless Pauli-Jordan distribution.
In QED the ghost fields do not couple and could therefore be eliminated.
We are, however, interested in a formulation of the gauge structure
which can be generalized to the nonabelian case, where
the ghosts seem to be indispensable [F,FP,DHKS1].

The only non-vanishing anticommutator of the spinor fields is
$$\{\psi (x),\psq (y)\}=-iS(x-y)=-i(i\gamma^\mu\d_\mu+m)
\Delta (x-y)\eqno(3.3)$$
and $\Delta$ denotes the massive Pauli-Jordan distribution.
The quantization of the photon field $A^\mu$ is
done in the Feynman gauge  with the commutator
$$ [A^\mu (x), A^\nu (y)]=ig^{\mu\nu}D (x-y).\eqno(3.4)$$
The $*$-operation is introduced by $A^{\mu\,*}= A^\mu$, 
$$ u^*\=d u,\quad\quad\quad\tilde u^*\=d -\tilde u.\eqno(3.5)$$
and $\psi(x)^{*}=\psq(x)\gamma_{0}$.
Unitarity {\bf (N2)} (with respect to the *-operation)
implies for the interacting fields
$$A^\mu_{{\rm int}\>{\cal L}}(x)^*=A^\mu_{{\rm int}\>{\cal L}}(x),\quad\quad
j^\mu_{{\rm int}\>{\cal L}}(x)^*=j^\mu_{{\rm int}\>{\cal L}}(x),\quad\quad
\psi_{{\rm int}\>{\cal L}}(x)^*\gamma_0=\psq_{{\rm int}\>
{\cal L}}(x),\eqno(3.6)$$
where $g$ is assumed to be real-valued and $j^\mu$ is the matter current
$$j^\mu (x)\=d :\psq (x)\gamma^\mu\psi (x):.\eqno(3.7)$$

In QED the field equations (2.16) read
$$\w A_{{\rm int}\>{\cal L}}^\mu (x)=-g(x)j_{{\rm int}\>{\cal L}}^\mu (x),
\eqno(3.8)$$
$$(i\gamma_\mu \d^\mu -m)\psi_{{\rm int}\>{\cal L}}(x)=
-g(x)(\gamma_\mu A^\mu \psi)_{{\rm int}\>{\cal L}}(x)\eqno(3.9)$$
and similar for $\psq_{{\rm int}\>{\cal L}}$.

An important restriction of the normalizations (i.e. the 
freedom in the extension
to the total diagonal) comes from the {\it Ward identities}
$${\bf (N5)}\quad\d_\mu^y T\Bigl( j^\mu (y)A_1(x_1)...A_n(x_n)\Bigr)=
i\sum_{j=1}^n\delta (y-x_j)
T\Bigl( A_1(x_1)...(\theta A_j)(x_j)...A_n(x_n)\Bigr),\eqno(3.10)$$
where $A_j$ is a sub Wick monom of $\cal L$ (3.1) and $(\theta A_j)\=d {d\over
d\alpha}\vert_{\alpha =0} A_{j\alpha}$ is the infinitesimal action of
the global $U(1)$ transformation
$$\psi\rightarrow \psi_\alpha\=d e^{i\alpha}\psi,\quad\quad\psq\rightarrow 
\psq_\alpha\=d e^{-i\alpha}\psq,
\quad\quad A^\mu\rightarrow A^\mu,\quad\quad u\rightarrow u,
\quad\quad \tilde u\rightarrow \tilde u,
\quad\quad\alpha\in{\bf R}.\eqno(3.11)$$
We prove in appendix B that all Ward
identities {\bf (N5)} can be fulfilled and are compatible with the
other above mentioned normalization conditions {\bf (N1)}, {\bf (N2)},
{\bf (N3)} and {\bf (N4)}. The subset of Ward identities 
involving only factors $j$ and
${\cal L}$
$$\d_\mu^y T\Bigl( j^\mu (y){\cal L}(x_1)...{\cal L}(x_k)j^{\mu_1}(x_{k+1})...
j^{\mu_{n-k}}(x_n)\Bigr) =0\eqno(3.12)$$
is equivalent to the free (perturbative) operator gauge invariance of QED
([S], [DHKS2] and section 5.1).
(3.10) implies the same Ward identities for the antichronological products 
$\bar T$ up to a global factor $(-1)$ on the r.h.s.. 
With that one easily derives from (3.10) the Ward
identities for the totally retarded products 
$$\d_\mu^y R\Bigl( A_1(x_1)...A_{n-1}(x_{n-1})j^\mu (y);A(x)\Bigr)=
i\delta (y-x)R\Bigl( A_1(x_1)...A_{n-1}(x_{n-1});(\theta A)(x)\Bigr)+$$
$$+i\sum_{k=1}^{n-1}\delta (y-x_k)
R\Bigl( A_1(x_1)...(\theta A_k)(x_k)...A_{n-1}(x_{n-1});A(x)\Bigr)$$
and
$$\d_\mu^y R\Bigl( A_1(x_1)...A_n(x_n);j^\mu (y)\Bigr)=
i\delta (y-x_k)R\Bigl( A_1(x_1)...\hat k...A_n(x_n);(\theta A_k)(x_k)\Bigr),$$
where $\hat k$ means that $A_k(x_k)$ is omitted. Especially we obtain
$$\d_\mu^x R({\cal L}(x_1)...{\cal L}(x_n);j^\mu (x))=0,\quad\quad
{\rm i.e.}\quad\quad\d_\mu j^\mu_{{\rm int}\>{\cal L}}(x)=0.\eqno(3.13)$$
Hence $j^\mu_{{\rm int}\>{\cal L}}$ is a conserved current.

The Ward identities {\bf (N5)} also imply that the 
corresponding charge operator
implements the infinitesimal $U(1)$-action 
$\theta$ on the interacting fields, i.e.
$$[j^0_{{\rm int}\>{\cal L}}(f), A_{{\rm int}\>{\cal L}}(x)]
=i(\theta A)_{{\rm int}\>{\cal L}}(x),$$
where $A$ is a sub Wick monom of $\cal L$ (3.1), and for the test function 
$f\in {\cal D}({\bf R}^4)$ we
assume that there exists $h\in {\cal D}({\bf R})$ such that
$$f(y)=h(y_0)\quad\forall y=(y_0,\vec y)\quad{\rm in}\>
{\rm a}\>{\rm neighbourhood}
\>{\rm of}\quad x+(\bar V_+\cup \bar V_-)\quad
{\rm and}\quad\int dy_0\,h(y_0)=1.$$
To prove this we first note
$[j^0_{{\rm int}\>{\cal L}}(f), A_{{\rm int}\>{\cal L}}(x)]=\int d^4y\,h(y_0)
[j^0_{{\rm int}\>{\cal L}}(y), A_{{\rm int}\>{\cal L}}(x)]$
by the support property (2.25) of the commutator, and by means 
of Proposition 2 this is equal to
$$-\sum_{n=0}^\infty {i^n\over n!}\int dy_1...dy_n\,g(y_1)...g(y_n)\int dy\,
\Bigl\{\Bigl([h(y_0)-h(y_0-a)]+h(y_0-a)\Bigr)\cdot$$
$$\cdot R({\cal L}(y_1)...{\cal L}(y_n)j^0 (y);A(x))-
\Bigl([h(y_0)-h(y_0-b)]+h(y_0-b)\Bigr) 
R({\cal L}(y_1)...{\cal L}(y_n)A(x);j^0 (y))\Bigr\}.$$
Due to  the support property (2.17) of the $R$-products we can choose $a$ and 
$b$ such that the contributions from $h(y_0-a)$ and $h(y_0-b)$ vanish. Setting
$k(y)\equiv k_0(y_0)\=d\int_{-\infty}^{y_0}dz\,[h(z)-h(z-a)]$, the 
$[h(y_0)-h(y_0-a)]$-term is equal to
$$...\int dy\,k(y)\d_\mu^y R({\cal L}(y_1)...{\cal L}(y_n)j^\mu (y);A(x))=
ik(x)(\theta A)_{{\rm int}\>{\cal L}}(x),$$
where we have inserted a Ward identity. For the $[h(y_0)-h(y_0-b)]$-term
we obtain $i(1-k(x))(\theta A)_{{\rm int}\>{\cal L}}(x)$ 
by a similar procedure.
So the sum of all terms is in fact $i(\theta A)_{{\rm int}\>{\cal L}}(x)$. $\w$

By means of the Ward identity (3.13) and Lemma 1, {\bf (A)} we obtain
$$\d_\mu A_{{\rm int}\>{\cal L}}^\mu (x)=\d_\mu A^\mu (x)
+\sum_{n=1}^{\infty}{i^{n+1}\over n!}\int d^4x_1...d^4x_n\,\cdot$$
$$\cdot\sum_{l=1}^ng(x_1)...\d_\mu g(x_l)...g(x_n)D^{\rm ret}(x-x_l)
R({\cal L}^0(x_1)...\hat l...{\cal L}^0(x_n);j^\mu (x_l)),\eqno(3.14)$$
where
$${\cal L}^0(x)\=d :\psq (x)\gamma_\mu A^\mu (x)\psi (x):,\quad\quad
{\rm i.e.}\quad\quad {\cal L}(x)=g(x){\cal L}^0(x).\eqno(3.15)$$
Thus in the formal partial adiabatic limit 
$g\vert_{x+\bar V_-}\rightarrow$ const.
the interacting field $\d_\mu A_{{\rm int}\>{\cal L}}^\mu (x)$ agrees
with the corresponding free one.

Let
$${\cal O}_{x,y}\=d (x+V_+)\cap (y+V_-).\eqno(3.16)$$
We will study the algebra $\tilde {\cal F}({\cal O})$ generated by
$$\{A_{{\rm int}\>{\cal L}}(f)=\int d^4x\,
A_{{\rm int}\>{\cal L}}(x)f(x)\vert f\in {\cal D(O)}, 
A=A^\mu,\psi,\psq,j^\mu...\},
\eqno(3.17)$$ 
where ${\cal O}$ is the double cone
$${\cal O}\=d {\cal O}_{(-r,\vec 0),(r,\vec 0)},\quad\quad (r>0)\eqno(3.18)$$
which is centered at the origin, and the test function
$g\in {\cal D}({\bf R}^4)$ in (3.1), (3.15) is assumed to fulfil
$$g(x)=e={\rm const.},\quad\quad\forall x\in {\cal O}.\eqno(3.19)$$

We are now going to compute some commutators of 
$\d_\mu A_{{\rm int}\>{\cal L}}^\mu (x)$ 
with elements of $\tilde {\cal F}({\cal O})$.

{\bf Proposition 3:} The following relations hold 
for the interacting fields at points $x,y\in 
{\cal O}$. 

$${\bf (1)}\quad\quad\quad
[\d_\mu A^\mu_{{\rm int}\>{\cal L}}(x),A^\nu_{{\rm int}\>{\cal L}}(y)]=
i\d^\nu D (x-y),\quad\quad\quad\quad\quad$$

$${\bf (2)}\quad\quad\quad
[\d_\mu A^\mu_{{\rm int}\>{\cal L}}(x),\d_\nu A^\nu_{{\rm int}\>
{\cal L}}(y)]=0,
\quad\quad\quad\quad\quad\quad\quad\quad\quad $$

$${\bf (3)}\quad\quad\quad
[\d_\mu A^\mu_{{\rm int}\>{\cal L}}(x),\psi_{{\rm int}\>{\cal L}}(y)]=
D (x-y)e\psi_{{\rm int}\>{\cal L}}(y),\quad$$

$${\bf (4)}\quad\quad\quad
[\d_\mu A^\mu_{{\rm int}\>{\cal L}}(x),\psq_{{\rm int}\>{\cal L}}(y)]=
-D (x-y)e\psq_{{\rm int}\>{\cal L}}(y),\>$$

$$\>{\bf (5)}\quad\quad\quad
[\d_\mu A^\mu_{{\rm int}\>{\cal L}}(x),{\cal L}^0_{{\rm int}\>{\cal L}}(y)]=
i(\d_\mu D)(x-y)j^\mu_{{\rm int}\>{\cal L}}(y).\quad$$

{\bf Proof:} Since $\d_\mu A^\mu_{{\rm int}\>{\cal L}}$ is a 
solution of the wave equation for $x\in {\cal O}$, 
it is sufficient to check these relations for the advanced or 
retarded part of the commutator. 

{\bf (1) and (2):} Inserting Lemma 1 {\bf (B)} and {\bf (C)} into 
Proposition 2 we obtain
$$[A^\mu_{{\rm int}\>{\cal L}}(x),A^\nu_{{\rm 
int}\>{\cal L}}(y)]_{\rm av}=-ig^{\mu\nu}
D^{\rm av}(x-y)-\sum_{n=1}^{\infty}{i^n\over n!}
\int d^4z_1...d^4z_n\,g(z_1)...g(z_n)$$
$$i\sum_{m=1}^nD^{\rm av}(x-z_m)R({\cal L}^0(z_1)...j^\mu(z_m)
...{\cal L}^0(z_n);A^\nu (y)).\eqno(3.20)$$

From the support properties of the retarded products (2.17)
we see that the integration over $z_m$
is confined to the double cone
$$z_m\in {\cal O}_{x,y}\subset {\cal O}.\eqno(3.21)$$ 
Let us now consider
$[\d_\mu A^\mu_{{\rm int}\>{\cal L}}(x),A^\nu_{{\rm int}\>{\cal L}}(y)]$.
We want to show that the divergence $\d_\mu^x$ of all terms in (3.20) with 
index $n\ge 1$ vanish.
In fact,  the divergence $\d_\mu^x$ can be written as
$-\d_\mu^{z_m}D^{\rm av}(x-z_m)$.
So formally, after a partial integration in $z_m$  we get two terms:
(i) A divergence of $R({\cal L}^0...j...{\cal L}^0;A)$ with respect to a
$j$-vertex, which vanishes due to the Ward identities {\bf (N5)}. (ii) A term
$\sim\d_\mu g(z_m)$ which vanishes 
since $g$ is constant within ${\cal O}$ (3.19,3.21).
The argument can easily 
be made rigorous  by smearing with test functions.
Hence 
$$[\d_\mu A^\mu_{{\rm int}\>{\cal L}}(x),A^\nu_{{\rm 
int}\>{\cal L}}(y)]_{\rm av}=-
i\d^\nu D^{\rm av} (x-y),\quad\quad\forall x,y\in {\cal O},\eqno(3.22)$$
which implies {\bf (1)}. Formula {\bf (2)} follows since $D$ is a solution 
of the wave equation.

{\bf (3):} To prove {\bf (3)} we proceed analogously to (3.20) and obtain
$$[\d_\mu A^\mu_{{\rm int}\>{\cal L}}(x),
\psi_{{\rm int}\>{\cal L}}(y)]_{\rm av}=
-\sum_{n=1}^{\infty}{i^n\over n!}
\int d^4z_1...d^4z_n\,g(z_1)...g(z_n)\cdot$$
$$\cdot i\sum_{m=1}^n
R({\cal L}^0(z_1)\ldots j^\mu (z_m)
\ldots{\cal L}^0(z_n);\psi(y))\d^x_\mu D^{\rm av}(x-z_m).\eqno(3.23)$$
We integrate by parts with respect to $z_m$
and insert the Ward identity
$$\d_\mu^{z_m}R\bigl({\cal L}^0(z_1)...j^\mu (z_m)
...{\cal L}^0(z_n);\psi(y)\bigr)=$$
$$=-\delta (z_m-y)R\bigl({\cal L}^0(z_1)...\hat m...
{\cal L}^0(z_n);\psi(y)\bigr).\eqno(3.24)$$
Because of $z_{m}\in {\cal O}$ (3.21)
the terms $\sim\d g(z_{m})$ vanish and we end up with
$$[\d_\mu A^\mu_{{\rm int}\>{\cal L}}(x),\psi_{{\rm 
int}\>{\cal L}}(y)]_{\rm av}=$$
$$=-D^{\rm av} (x-y)g(y)\psi_{{\rm int}\>{\cal L}}(y)
\quad\quad {\rm for}\quad\quad
x,y\in {\cal O},\eqno(3.25)$$ 
This proves {\bf (3)}. 

{\bf (4):} The relation for the conjugate spinor follows by applying
the *-operation. 

{\bf (5):} By means of Proposition 2 we get
$$[\d_\mu A^\mu_{{\rm int}\>{\cal L}}(x),
{\cal L}^0_{{\rm int}\>{\cal L}}(y)]_{\rm ret}=$$
$$\sum_{n=0}^{\infty}{i^n\over n!}
\int d^4z_1...d^4z_n\,g(z_1)...g(z_n)
\d_\mu^x R({\cal L}^0(z_1)...{\cal L}^0(z_n){\cal L}^0(y);A^\mu(x)).
\eqno(3.26)$$
From part {\bf (A)}  of Lemma 1 this is equal to
$$\sum_{n=0}^{\infty}{i^{n+1}\over n!}\int dz_1...dz_n\,g(z_1)...g(z_n)
\{\d_\mu D^{\rm ret}(x-y)R(({\cal L}^0(z_1)...{\cal L}^0(z_n);j^\mu(y))+$$
$$+\sum_{m=1}^n\d_\mu D^{\rm ret}(x-z_m)R({\cal L}^0(z_1)...\hat 
m...{\cal L}^0(z_n){\cal L}^0(y);j^\mu (z_m))\}.\eqno(3.27)$$
By a partial integration with respect to $z_m$ and the 
same reasoning as above, we
see that the second term does not contribute, hence we find
$$[\d_\mu A^\mu_{{\rm int}\>{\cal L}}(x),
{\cal L}^0_{{\rm int}\>{\cal L}}(y)]_{\rm ret}=
i(\d_\mu D^{\rm ret})(x-y)j^\mu_{{\rm int}\>{\cal L}}(y),\eqno(3.28)$$
which finally implies {\bf (5)}. $\quad\quad\w$
\vskip 1cm
{\trm 4. Connection of observable algebras and field algebras in
perturbative gauge theories}
\vskip 1cm
The observation (2.1) applies to the algebra $\tilde {\cal F}({\cal O})$ (3.17)
with a unitary formal power series $V$. We conclude that the net of 
algebras of local interacting fields $\tilde {\cal F}({\cal O}_1),\>{\cal O}_1
\subset {\cal O}$, up to unitary equivalence is uniquely determined by
$g\vert_{\cal O}$. Since ${\cal O}$ is arbitrary, the full net of local 
algebras can be constructed without performing the adiabatic limit 
$g\rightarrow {\rm constant}$.

This general procedure can be applied also to {\it gauge theories}. 
But there the (local)
algebras of interacting fields contain unphysical fields like vector
potentials and ghosts. Therefore the question arises whether 
there is a local construction of the algebra of 
observables and the physical Hilbert space, i.e. a (pre)Hilbert space on 
which the observables can be faithfully represented.
\vskip 0.5cm
{\it 4.1 Local construction of observables in gauge theories}
\vskip 0.5cm
Let ${\cal O}\rightarrow {\cal F}({\cal O})$ be a net of ${\bf Z}_2$-graded
*-algebras. In addition, we are 
given a graded derivation $s$ on $\cup_{\cal O}{\cal F}({\cal O})
=:{\cal F}$ with $s^2=0$, $s({\cal F}({\cal O}))\subset {\cal F}({\cal O})$
and $s(F^*)=-(-1)^{\delta (F)}s(F)^*$, where $(-1)^{\delta (F)}$ is the
${\bf Z}_2$-gradiation of $F\in {\cal F}$.

The observables should be $s$-invariant. Therefore, we consider the kernel of
$s$, ${\cal A}_0:=s^{-1}(0)\subset {\cal F}$, and ${\cal A}_0
({\cal O}):= {\cal A}_0\cap {\cal F}({\cal O})$. ${\cal A}_0$ is a *-algebra:
$$A,B\in {\cal A}_0\Longrightarrow s(AB)=s(A)B+(-1)^{\delta (A)}As(B)=0,
\quad\quad\quad {\rm i.e.}\quad AB\in {\cal A}_0.\eqno(4.1)$$
We set ${\cal A}_{00}:=s({\cal F})$. Due to $s^2=0$ it is a subspace of
${\cal A}_0$; it is even a 2-sided ideal in ${\cal A}_0$
$$s(F)A=s(FA)-(-1)^{\delta (F)}Fs(A)=s(FA)\in {\cal A}_{00}\eqno(4.2)$$
and similarly $As(F)\in {\cal A}_{00}$ if $A\in {\cal A}_0$. 
The algebra of observables  is defined as the quotient
$${\cal A}\=d {{\cal A}_0\over {\cal A}_{00}}\eqno(4.3)$$
and 
$${\cal O}\rightarrow {\cal A}({\cal O})\=d {{\cal A}_0\cap {\cal F}({\cal O})
\over {\cal A}_{00}\cap {\cal F}({\cal O})}\eqno(4.4)$$
is the net of algebras of local observables.
\vskip 0.5cm
{\it 4.2 Representation of the observables in the physical pre Hilbert space}
\vskip 0.5cm
We now ask: under which conditions does ${\cal A}$ 
have a nontrivial representation
by operators on a pre Hilbert space, such that
$$(A^*\phi,\psi)=(\phi,A\psi),\quad\quad\quad\forall A\in{\cal A}.\eqno(4.5)$$
For this purpose we assume that ${\cal F}$ has a faithful representation on 
an inner product space $({\cal K},<.,.>)$ such that
$$<F^*\phi,\psi>=<\phi,F\psi>,\quad\quad\quad\forall F\in {\cal F},$$
and $s$ is implemented by an operator $Q$ on ${\cal K}$, i.e.
$$s(F)=QF-(-1)^{\delta (F)}FQ,\eqno(4.6)$$
such that
$$<Q\phi,\psi>=<\phi,Q\psi>\quad\quad\quad {\rm and}\quad\quad\quad Q^2=0.
\eqno(4.7)$$
The assumptions (4.7) are made in order to fulfil $s(F^*)=
-(-1)^{\delta (F)}s(F)^*$ and $s^2=0$.
Note that if the inner product on ${\cal K}$ is positive definite, we find
$<Q\phi,Q\phi>=<\phi,Q^2\phi>=0$, hence $Q=0$ and thus also $s=0$. Hence
for nontrivial $s$ the inner product must necessarily be indefinite.

Since the physical states should be $s$-invariant, we consider the kernel 
of $Q$: ${\cal K}_0\=d {\rm Ke}\,Q$. Let ${\cal K}_{00}$ 
be the range of $Q$.
Because of $Q^2=0$ we have ${\cal K}_{00}\subset {\cal K}_0$. We assume:
$${\bf (Positivity)}\quad\quad\quad 
{\rm (i)}\quad <\phi,\phi>\geq 0\quad\quad\forall 
\phi\in {\cal K}_0,$$
$$\quad\quad\quad {\rm and}\quad\quad {\rm (ii)}\quad\phi\in {\cal K}_0\quad 
\wedge\quad <\phi,\phi>=0\quad\Longrightarrow
\quad \phi\in {\cal K}_{00}.\eqno(4.8)$$

Then 
$${\cal H}\=d {{\cal K}_0\over {\cal K}_{00}},\quad\quad\quad
<[\phi_1],[\phi_2]>_{\cal H}:\=d <\psi_1,\psi_2>_{\cal K},
\quad\psi_j\in [\phi_j]:=\phi_j+{\cal K}_{00}\eqno(4.9)$$
is a pre Hilbert space. (Due to (4.7) the definition of $<[\phi_1],
[\phi_2]>_{\cal H}$ is independent of the choice of the representatives 
$\psi_j\in [\phi_j],\>j=1,2$.)

Now we assure that 
$$\pi ([A])[\phi]\=d [A\phi]\eqno(4.10)$$ 
is a well defined
representation on ${\cal H}$ (where $A\in {\cal A}_0,\,\phi\in 
{\cal K}_0,\,[A]:=A+{\cal A}_{00}$). Namely, let $A+s(B),\> A\in {\cal A}_0,\>
B\in {\cal F}$, be a 
representative of $[A]\in {\cal A}$ in ${\cal F}$, and let
$\phi+Q\psi,\>\phi\in {\cal K}_0,\>\psi\in {\cal K}$ be a representative of 
$[\phi]\in {\cal H}$ in ${\cal K}$. We have to show that $A\phi\in
{\cal K}_0$ and $(A+s(B))(\phi+Q\psi)-A\phi\in {\cal K}_{00}=Q{\cal K}$.
But $QA\phi=s(A)\phi+(-1)^{\delta (A)}AQ\phi=0$, and
$$s(B)\phi+(A+s(B))Q\psi = (QB-(-1)^{\delta (B)}BQ)(\phi+Q\psi)
-(-1)^{\delta (A)}(s(A)\psi-QA\psi)=$$
$$=QB(\phi+Q\psi)+(-1)^{\delta (A)}QA\psi\in Q{\cal K}.$$
\vfill
\eject
{\it 4.3 Stability under deformations}
\vskip 0.5cm
It is gratifying that the described structure is {\it stable under 
deformations}, e.g. by turning on the interaction.
Let ${\cal K}$ be fixed and replace $F\in {\cal F}$ by a formal power series
$\tilde F=\sum_n g^nF_n$ with $F_0=F$ and $F_n\in {\cal F},\,
\delta(F_{n})={\rm const}$. In the same
way replace $s$ and $Q$ by formal power series $\tilde s=\sum_n g^ns_n$
(each $s_n$ is a graded derivation), $\tilde Q=\sum_n g^nQ_n$, 
$Q_n\in L({\cal K})$, with $s_0=s$, $Q_0=Q$ and
$$\tilde s^2=0,\quad\tilde Q^2=0,\quad <\tilde Q\phi,\psi>=<\phi,
\tilde Q\psi>\quad {\rm and}\quad \tilde s(\tilde F)=\tilde Q\tilde F
-(-1)^{\delta (\tilde F)}\tilde F\tilde Q.\eqno(4.11)$$
We can then define $\tilde {\cal A}\=d {{\rm Ke}\,\tilde s\over 
{\rm Ra}\,\tilde s}$. ${\cal K}_0$ and ${\cal K}_{00}$
have to be replaced by formal power series $\tilde {\cal K}_0:=
{\rm Ke}\,\tilde Q$ and $\tilde {\cal K}_{00}:={\rm Ra}\,\tilde Q$
with coefficients in ${\cal K}$. Due to the above result, the algebra
$\tilde {\cal A}$ has a natural representation on 
$\tilde {\cal H}\=d {\tilde {\cal K}_0\over \tilde {\cal K}_{00}}$.
The inner product on ${\cal K}$ induces an inner product 
on $\tilde{\cal H}$ which assumes values in the formal 
power series over ${\bf C}$. 
We adopt the point of view that a formal power series $\tilde b=\sum_ng^nb_n,
\>b_n\in {\bf C}$ is {\it positive} if there is another formal power series
$\tilde c=\sum_ng^nc_n,\>c_n\in {\bf C}$ with $\tilde c^*\tilde c=\tilde b$,
i.e. $b_n=\sum_{k=0}^n \bar c_k c_{n-k}$. This is equivalent to the condition
\footnote{$^4$}{Bordemann and Waldmann [BW] work with a weaker
definition of positivity in the case
of a formal Laurent series with real coefficients: they only 
require that the smallest non-vanishing coefficient is positive, it does
not need to be an even coefficient.}

$$b_n\in {\bf R}\quad\quad\quad\forall n\in {\bf N}_0,\eqno(4.12)$$
and
$$\exists k\in {\bf N}_0\cup\{\infty\}\quad\quad {\rm such}\quad {\rm that}
\quad\quad b_l=0\quad\forall l<2k,$$
$$\quad\quad {\rm and}\quad\quad b_{2k}>0\quad\quad {\rm in}\quad {\rm the}
\quad {\rm case}\quad k<\infty.\eqno(4.13)$$

We now show that the assumptions concerning the positivity of the inner
product are automatically
fulfilled for the deformed theory, if they hold true in the undeformed model.

{\bf Theorem 4:} Let the positivity assumption (4.8) 
be fulfilled in zeroth order. Then

(i) $<\tilde\phi,\tilde\phi>\geq 0\quad\quad\forall 
\tilde\phi\in \tilde {\cal K}_0$,

(ii) $\tilde\phi\in \tilde {\cal K}_0\quad \wedge\quad <\tilde\phi,
\tilde\phi>=0\quad\Longrightarrow\quad \tilde\phi\in \tilde {\cal K}_{00}.$

(iii) For every $\phi\in {\cal K}_0$
there exists a power series $\tilde\phi\in\tilde {\cal K}_0$ with 
$(\tilde\phi)_0=\phi$.

(iv) Let $\pi$ and $\tilde\pi$ be the representations 
(4.10) of ${\cal A},\,\tilde
{\cal A}$ on ${\cal H},\,\tilde {\cal H}$ respectively. Then

$\tilde{\pi}(\tilde { A})\neq 0$ if $\pi(A_{0})\neq 0$.

{\it Proof of Theorem 4}, (i) and (ii): 
Let $\tilde\phi\in \tilde {\cal K}_0$ and
$b_n=\sum_{k=0}^n<\phi_k,\phi_{n-k}>$. $b_n$ clearly is real. $\tilde Q
\tilde \phi =0$ implies $Q_0\phi_0=0$, hence $\phi_0\in {\cal K}_0$
and $b_0\geq 0$. If $b_0>0$ (i) follows. If $b_0=0$ we know that there is some 
$\psi_0\in {\cal K}$ with $\phi_0=Q_0\psi_0$. Let $\psi^{(0)}_k:=
\psi_0\delta_{k,0}$ and $\tilde\psi^{(0)}:=\sum_kg^k\psi^{(0)}_k=\psi_0$. Then
$$\tilde\eta^{(0)}:=\tilde\phi -\tilde Q\tilde\psi^{(0)},\eqno(4.14)$$
is a formal power series with vanishing term of zeroth order. We now proceed 
by induction and assume that $b_0=b_1=...=b_{2n}=0$ and there is some formal 
power series $\tilde\psi^{(n)}=\sum_kg^k\psi^{(n)}_k$ with coefficients
in ${\cal K}$ such that
$$\tilde\eta^{(n)}:=\sum_kg^k\eta^{(n)}_k=
\tilde\phi -\tilde Q\tilde\psi^{(n)}\eqno(4.15)$$
vanishes up to order $n$. Then
$$b_{2n+1}=<\tilde\eta^{(n)}+\tilde Q\tilde\psi^{(n)},
\tilde\eta^{(n)}+\tilde Q\tilde\psi^{(n)}>_{2n+1}=<\tilde\eta^{(n)},
\tilde\eta^{(n)}>_{2n+1}=0\eqno(4.16)$$
and $b_{2n+2}=<\eta^{(n)}_{n+1},\eta^{(n)}_{n+1}>$. Since
$\tilde Q\tilde \eta^{(n)} =0$ we get $Q_0\eta^{(n)}_{n+1}=0$, i.e.
$\eta^{(n)}_{n+1}\in {\cal K}_0$ and $b_{2n+2}\geq 0$. If $b_{2n+2}> 0$
we obtain (i), otherwise $\exists \psi_{n+1}\in {\cal K}$ with
$\eta^{(n)}_{n+1}=Q_0\psi_{n+1}$, and we can define
$$\psi^{(n+1)}_k:=\psi^{(n)}_k+\delta_{n+1,k}\psi_{n+1}.
\eqno(4.17)$$
One easily verifies
$$(\tilde\phi -\tilde Q\tilde\psi^{(n+1)})_k=0,\quad\quad
\forall k=0,1,...,n+1.\eqno(4.18)$$
Either the induction stops at some $n$ or we find a $\tilde \psi$ with
$\tilde\phi =\tilde Q\tilde\psi$, i.e. $\tilde\phi\in \tilde {\cal K}_{00}$.

To prove (iii) we again proceed by induction and assume that there exists a 
power series $\tilde\phi^{(n)}$ such that $\tilde Q\tilde\phi^{(n)}$ vanishes
up to order $n$. This is certainly true for $n=0$. Then
$0=(\tilde Q^2\tilde \phi^{(n)})_{n+1}=Q_0(\tilde Q\tilde\phi^{(n)})_{n+1}$, 
hence $(\tilde Q\tilde\phi^{(n)})_{n+1}\in {\cal K}_0$. Moreover $0=\break
<\tilde Q\tilde\phi^{(n)},\tilde Q\tilde\phi^{(n)}>_{2n+2}=<(\tilde Q
\tilde\phi^{(n)})_{n+1},(\tilde Q\tilde\phi^{(n)})_{n+1}>$, thus
$(\tilde Q\tilde\phi^{(n)})_{n+1}\in {\cal K}_{00}$ and there exists a 
$\phi_{n+1}\in
{\cal K}$ with $(\tilde Q\tilde\phi^{(n)})_{n+1}+Q_0\phi_{n+1}=0.$ We then set
$(\tilde\phi^{(n+1)})_k:=(\tilde\phi^{(n)})_k+\delta_{n+1,k}\phi_{n+1}$
and find that $\tilde Q\tilde\phi^{(n+1)}$ vanishes up to order $n+1$.
$$\tilde\phi:=\lim_{n\to\infty}\tilde\phi^{(n)}\in\tilde {\cal K}_0
\eqno(4.19)$$ 
is then the wanted formal power series.

It remains to prove (iv). $\tilde{\pi}(\tilde{A})=0$ means  
$\tilde A=\sum_kg^kA_k\in {\rm Ke}\,\tilde s$ 
with $\tilde A\tilde\phi\in 
\tilde {\cal K}_{00},\>\forall \tilde\phi\in \tilde {\cal K}_0$.
By means of (iii) this implies in zeroth order $A_{0}\phi_{0}\in {\cal K}_{00},
\>\forall \phi_{0}\in {\cal K}_{0}$, i.e. $\pi(A_{0})=0$.$\quad\w$ 

Note that $\phi\rightarrow\tilde\phi$ is non-unique and this
holds true also for the induced relation between ${\cal 
H}$ and $\tilde {\cal H}$.

The unit $\tilde 1$ in an algebra of formal power series is
$\tilde 1=(1,0,0,....)=1g^0$, and $\tilde a=\sum _{k=0}^\infty
a_k g^k$ is invertible iff $a_0$ is invertible.
We denote by $\tilde {\bf C}$ the formal power series over ${\bf C}$ and
consider $\tilde {\cal K}\=d\{\tilde\phi=\sum_n\phi_ng^n|\phi_n\in 
{\cal K}\}$ and $\tilde {\cal F}\=d\{\tilde F=\sum_n F_ng^n|F_n\in 
{\cal F}\}$ as $\tilde {\bf C}$-modules. This is possible because the usual 
multiplication of power series yields maps
$$\tilde {\bf C}\times \tilde {\cal F}\rightarrow \tilde {\cal F}:
(\tilde a,\tilde A)\rightarrow \tilde a\tilde A=\tilde A\tilde a,\quad\quad
\quad\tilde {\bf C}\times \tilde {\cal K}\rightarrow \tilde {\cal K}:
(\tilde a,\tilde \phi)\rightarrow \tilde a\tilde\phi=\tilde\phi\tilde a,$$
which fulfil the relations
$$\tilde A(\tilde a\tilde\phi)=\tilde a(\tilde A\tilde\phi)=
(\tilde a\tilde A)\tilde\phi,\quad\quad (\tilde a\tilde A)^*=\tilde a^*
\tilde A^*,\quad\quad <\tilde a\tilde\phi,\tilde b\tilde\psi>=
\tilde a^*\tilde b<\tilde\phi,\tilde\psi>\eqno(4.20)$$
and
$$\tilde s(\tilde a\tilde A)=\tilde a\tilde s(\tilde A).\eqno(4.21)$$
Also the physical pre Hilbert space $\tilde {\cal H}$ and the algebra 
of observables $\tilde {\cal A}({\cal O})$ are 
$\tilde {\bf C}$-modules, and the multiplications by a 'scalar'
$$\tilde {\bf C}\times \tilde {\cal A}({\cal O})\rightarrow \tilde 
{\cal A}({\cal O}):(\tilde a,[\tilde A])\rightarrow \tilde a[\tilde A]=
[\tilde a\tilde A]=[\tilde A]\tilde a\eqno(4.22)$$
$$\tilde {\bf C}\times \tilde {\cal H}\rightarrow \tilde {\cal H}:
(\tilde a,[\tilde \phi])\rightarrow \tilde a[\tilde\phi]=[\tilde a\tilde\phi]
=[\tilde\phi]\tilde a\eqno(4.23)$$
satisfy (4.20). We are now going to prove that every $[\tilde\phi]\in\tilde 
{\cal H}$ can be normalized:

{\bf Corollary 5}:  For every $[\tilde\phi]\in \tilde {\cal H},\>
[\tilde\phi]\not= 0$, there exist $[\tilde\psi]\in \tilde {\cal H}$ and
$\tilde a\in \tilde {\bf C}$ such that 
$$[\tilde\phi]=\tilde a[\tilde\psi]\quad\quad {\rm and}
\quad\quad <[\tilde\psi],[\tilde\psi]>=1.\eqno(4.24)$$

{\it Proof}: We set $\tilde b:=<[\tilde\phi],[\tilde\phi]>
\in \tilde {\bf C}$. From Theorem 4 we know $\tilde b=\sum_{n=2k}^\infty 
b_ng^n,\>b_n\in {\bf R},\>b_{2k}>0$.

{\it Case (1)}, $k=0$: There exists an invertible $\tilde a\in \tilde {\bf C}$
with $\tilde a^* \tilde a=\tilde b$. Then $[\tilde\psi]:=\tilde a^{-1}
[\tilde\phi]$ satisfies the assertion (4.24).

{\it Case (2)}, $k>0$: We consider a representative $\tilde\phi=\sum_n
\phi_ng^n$ of $[\tilde\phi]$. Due to $<\phi_0,\phi_0>=b_0=0$ and 
$Q_0\phi_0=0$, there exists $\eta_0\in {\cal K}$ with $Q_0\eta_0=\phi_0$.
Then we can define $\tilde\tau_1$ by $g\tilde\tau_1:=\tilde\phi-
\tilde Q\eta_0$ which fulfills $\tilde\tau_1\in\tilde {\cal K}_0$ and
$[\tilde\phi]=g[\tilde\tau_1]$. Hence $<[\tilde\tau_1],[\tilde\tau_1]>
=g^{-2}\tilde b$. If $k>1$ we repeat this procedure 
(starting with $[\tilde\tau_1]$ instead of $[\tilde\phi]$) until we obtain
a $\tilde\tau_k\in\tilde {\cal K}_0$ with $[\tilde\phi]=g^k[\tilde\tau_k]$ 
and hence $<[\tilde\tau_k],[\tilde\tau_k]>=g^{-2k}\tilde b$. Similarly
to case (1) we conclude that there exists an invertible 
$\tilde c\in \tilde {\bf C}$ with $\tilde c^* \tilde c=g^{-2k}\tilde b$.
Then $[\tilde\psi]:=\tilde c^{-1}[\tilde\tau_k]$ satisfies  
$<[\tilde\psi],[\tilde\psi]>=1$ and $[\tilde\phi]=g^k\tilde c[\tilde\psi]$,
i.e. (4.24) is satisfied for $\tilde a:=g^k\tilde c$. $\quad\quad\w$
\vskip0.5cm
A state $\omega$ on the algebra of observables $\tilde {\cal A}({\cal O})$
is defined by

(i) $\omega$: $\tilde {\cal A}({\cal O})\rightarrow \tilde {\bf C}$ is
linear, i.e. $\omega (\tilde a[\tilde A]+[\tilde B])=
\tilde a\omega ([\tilde A])+\omega ([\tilde B])$,

(ii) $\omega ([\tilde A]^*)=\omega ([\tilde A])^*\quad\quad
\forall [\tilde A]\in\tilde {\cal A}({\cal O})$,

(iii) $\omega ([\tilde A]^*[\tilde A])\geq 0\quad\quad
\forall [\tilde A]\in\tilde {\cal A}({\cal O})\quad\quad$ and

(iv) $\omega (\tilde {\bf 1})=\tilde 1.\quad\quad\quad\quad\quad\quad
\quad\quad\quad\quad\quad\quad\quad\quad\quad\quad\quad\quad\quad
\quad\quad\quad\quad\quad\quad\quad\quad\quad\quad\quad$\break
The constructed physical states, i.e. the vector states 
$$\omega_{\tilde\phi}([\tilde A])=<[\tilde\phi],[\tilde A][\tilde\phi]>,
\quad\quad [\tilde\phi]\in\tilde {\cal H},\eqno(4.25)$$
satisfy obviously (i), (ii) and, if $<[\tilde\phi],[\tilde\phi]>=1$,
also (iv). The positivity (iii) of the states $\omega_{\tilde\phi}$
is ensured by

{\bf Corollary 6} (Positivity of the Wightman distributions of gauge 
invariant fields): Let the algebra $\tilde {\cal A}$ be generated by the
$\tilde s$-invariant fields $\tilde\phi_1,...,\tilde\phi_l$ and let
$$\tilde A:=\sum_{n=0}^k\int\sum_{j_1,...,j_n}f_{j_1...j_n}(x_1,...,x_n)
\tilde\phi_{j_1}(x_1)...\tilde\phi_{j_n}(x_n)dx_1...dx_n,\quad\quad
f_{j_1...j_n}\in {\cal D}({\bf R}^{4n}),\eqno(4.26)$$
and $\tilde\phi\in \tilde {\cal K}_0$. Then 
$$<\tilde\phi,\tilde A^*\tilde A\tilde\phi>
\geq 0.\eqno(4.27)$$

{\it Proof}: Note $\tilde A\tilde\phi\in\tilde {\cal K}_0$ and apply
part (i) of Theorem 4. $\quad\quad\w$

{\it Remark}: Let us assume $\tilde Q=Q_0$. This situation occurs if 
the adiabatic limit exists ([KO], see also sect. 5.2), e.g. in massive 
gauge theories. Then $\tilde Q\tilde\phi=0$ means $Q_0\phi_k=0,\>\forall k$.
Therefore, in this case the physical pre Hilbert space $\tilde {\cal H}$
is the space of formal power series with coefficients in ${\cal H}$,
$$\tilde {\cal H}=\tilde {\bf C}{\cal H}\quad\quad\quad ({\rm if}\quad
\tilde Q=Q_0).\eqno(4.28)$$
But the states $\omega_\phi$ on $\tilde {\cal A}({\cal O})$ induced by vectors
$\phi\in {\cal H}$ remain $\tilde {\bf C}$-valued functionals.
\vfill
\eject
{\trm 5. Verification of the assumptions in the example of QED}
\vskip 1cm
The construction in the previous section relies on some assumptions, which we
are now going to verify in QED. The deformation is given by going over from 
the free theory to the interacting fields discussed in sections 2 and 3.
For the free and the interacting theory we will first define the 
BRST-transformation $s$ ($\tilde s$ resp.) and then we will 
construct a nilpotent 
and hermitean operator $Q$ ($\tilde Q$) which implements $s$ ($\tilde s$) in a 
representation space with indefinite inner product. Then the local observables
(defined by (4.4)) are naturally represented on ${\cal H}={{\rm Ke}\,Q\over
{\rm Ra}\,Q}$ ($\tilde {\cal H}={{\rm Ke}\,\tilde Q\over {\rm Ra}\,\tilde Q}$)
by (4.10). It remains to prove the positivity of the inner product induced in
${\cal H}$ ($\tilde {\cal H}$). For the free theory we will do this by giving 
explicitly (distinguished) representatives of the equivalence classes in 
${\cal H}$. Then we conclude from Theorem 4 that positivity holds true also for
$\tilde {\cal H}$.
\vskip 0.5cm
{\it 5.1 The free theory}
\vskip 0.5cm
We consider the field algebra ${\cal F}$ which
is generated by the free fields $A^\mu,\,\psi,\,
\psq,\,u,\,\tilde u$, the Wick monomials $j^\mu =:\psq\gamma^\mu\psi:,\,
\gamma_\mu A^\mu\psi,\,
\psq\gamma_\mu A^\mu,\,{\cal L}^0=j_\mu A^\mu$ and the 
derivatives of free fields 
$\d_\mu A^\mu$,
$F^{\mu\nu}=\d^\mu A^\nu-\d^\nu A^\mu$. This algebra has a faithful 
representation on the Fock space ${\cal K}={\cal K_A}\otimes {\cal K_\psi}
\otimes {\cal K}_g$ of free fields (appendix A).
The ${\bf Z}_2$-gradiation is $(-1)^{\delta (F)}$, where $F$ is a monomial
in ${\cal F}$ and ${\delta (F)}$ is the ghost number
$$[Q_u,F]=\delta (F)F,\quad\quad\quad\quad Q_u\=d 
i\int_{x_0={\rm const.}}d^3x:\tilde u(x){\dl}_0 u(x):.\eqno(5.1)$$
Note $\delta (u)=-1,\>\delta (\tilde u)=1$.  
The graded derivation $s$ is the 
BRST-transformation of free fields
$$s(A^\mu)=i\d^\mu u,\quad s(\psi)=0,\quad s(\psq)=0,\quad s(u)=0,
\quad s(\tilde u)=-i\d_\mu A^\mu.\eqno(5.2)$$
The transformation of Wick monomials and derivated free fields is given by
$$s(:\phi_1(x)\phi_2(x)...:)=:s(\phi_1)(x)\phi_2(x)...:+(-1)^{\delta (\phi_1)}
:\phi_1(x)s(\phi_2)(x)...:+...\eqno (5.3)$$
and by translation invariance of $s$
$$s(\d_\mu\phi(x))=\d_\mu^x s(\phi)(x).\eqno(5.4)$$
This transformation is implemented by the free Kugo-Ojima charge [DHKS1]
$$Q\=d \int_{x_0={\rm const.}} d^3x\,(\d_{\nu}A^{\nu}(x)){\dl}_0u(x),
\eqno(5.5)$$
which fulfills $Q^*=Q$,
\footnote{$^5$}{We restrict all operators 
(resp. formal power series of operators)
to the dense invariant domain ${\cal D}$ and, therefore, there is no difference
between symmetric and self-adjoint operators.}
$[Q_u,Q]=-Q$ and $Q^2=0$. 
This is verified in appendix A. 
Moreover, it is shown there that the inner product
$<.,.>$ is positive semidefinite on ${\rm Ke}\,Q$ and that the 
space of null vectors
in ${\rm Ke}\,Q$ is precisely ${\rm Ra}\,Q$ [DHS1].

The existence of the integral in (5.5) can be proven by the following
method due to Requardt [R]. We
smear out $J^0=\d_{\nu}A^{\nu}{\dl}_0u$ with 
$k(x_0)h({\bf x})\in {\cal D}({\bf R}^4)$,
where $\int dx_0\,k(x_0)=1$ and $h$ is a smeared characteristic function of
$\{{\bf x}\in {\bf R}^3, |{\bf x}|\leq R\}$ for some $R>0$. By scaling
$k_\lambda (x_0):=\lambda k(\lambda x_0),\>h_\lambda 
({\bf x}):=h(\lambda {\bf x}),\>
Q_\lambda :=\int d^4x\,k_\lambda (x_0)h_\lambda ({\bf x})J^0(x)$ 
one easily finds
$\lim_{\lambda \to 0}\| Q_\lambda\Omega\| =0$ w.r.t. a suitable Krein
space norm (cf Appendix A). In addition, due to current
conservation, $\lim_{\lambda \to 0}[Q_\lambda,\phi(y)]_\mp =\int d^3x\,[J^0(x),
\phi(y)]_\mp$ (note that the latter integral 
exists since the region of integration is 
bounded) for $R$ big enough compared to the support of $k$. 
Therefore, the strong limit
$\lim_{\lambda \to 0} Q_\lambda$ exists on a 
dense subspace and agrees with (5.5).

Unfortunately, the representation (4.10) of the 
observables ${\cal A}={{\rm Ke}\,s\over 
{\rm Ra}\,s}$ on the physical pre Hilbert space
${\cal H}={{\rm Ke}\,Q\over {\rm Ra}\,Q}$ is {\it not faithful .}
The counterexample is $[u(f)],\>f\in {\cal D}({\bf R^4})$ real-valued, 
which induces a non-trivial element of ${\cal A}$ if $\int f 
d^4x\neq 0$.
Namely, due to $u(f)^*u(f)=u(f)^2=0$, it is represented by zero on ${\cal H}$.
(This holds true for each representation in which $<\cdot,\cdot >$ is positive
definite.) Since $u(\partial_{\mu}h)=i\,s(A_{\mu}(h)),\> 
h\in{\cal D}({\bf R^4})$,
${\cal A}$ has the following structure:
$${\cal A}={\cal A}^{(0)}\oplus u_{0}{\cal A}^{(0)},\eqno(5.6)$$
where $u_{0}$ is the rest class of $u(f)$, $\int f d^4x=1$ 
and where ${\cal A}^{(0)}$ is the subalgebra with ghost 
number zero.
 
The representation (4.10) of ${\cal A}^{(0)}$ on ${\cal H}$ is faithful. 
To make this 
plausible we mention that ${\cal A}^{(0)}$ is generated by $[F^{\mu\nu}],
\,[\psi],\,[\psq]$ and Wick monomials thereof, whereas the 'canonical' 
representatives of ${\cal H}$ are the states containing transversal photons, 
electrons and positrons only (A.39).

The interaction Lagrangian of QED is $s$-invariant up to a 
divergence of a local field,
$$[Q,{\cal L}^0(x)]=i\d_\mu {\cal L}^{1\,\mu}(x),\quad\quad\quad
{\cal L}^{1\,\mu}\=d :\psq\gamma^\mu\psi:u.\eqno(5.7)$$
Thus in the formal adiabatic limit the integral of the 
Lagrangian becomes invariant.
In  [DHKS1,DHKS2] the following Ward identities were 
postulated: 
$$[Q,T({\cal L}^0(x_1)...{\cal L}^0(x_n))]=\sum_{l=1}^n \d_\mu^{x_l}
T({\cal L}^0(x_1)...{\cal L}^{1\,\mu}(x_l)...{\cal L}^0(x_n)).\eqno(5.8)$$
('free (perturbative) operator gauge invariance', compare (3.12)). Provided 
the adiabatic limit exists this condition implies the 
$s$-invariance of the S-matrix; hence the S-matrix induces a 
unitary operator on the physical Hilbert space ${\cal H}=
{{\rm Ke}\,Q\over {\rm Ra}\,Q}$ [DHS1,K].

The nice feature of the condition (5.8) is that its formulation 
makes sense independent of the adiabatic limit.
So, if the normalizations are suitably chosen, 
the free (perturbative) operator gauge invariance (5.8) (more precisely
the corresponding C-number identities which imply (5.8)) 
could been proven to hold to
all orders in QED [DHS2,S] and
also in $SU(N)$ Yang-Mills theories [DHS1,D1] and to  imply 
(in the latter case) the usual Slavnov-Taylor identities [D2].
In addition, it  determines to a large extent the 
possible structure of the model. Stora [St1] found that making 
a general ansatz for
the interaction Lagrangian of selfinteracting gauge fields, the Ward identities
(5.8) require the coupling parameters to be totally antisymmetric structure
constants of a Lie group. Moreover, (5.8) was used for a 
derivation of all the couplings of the standard model of 
electroweak interactions (especially the Higgs potential) [DS].
 
We emphasize that (5.8) is a {\it pure quantum 
formulation of gauge invariance}, without reference to classical physics.
\vfill\eject
{\it 5.2 The interacting theory: 
construction of the interacting Kugo-Ojima charge}
\vskip 0.5cm
We now replace the free fields (including  Wick monomials and 
derivatives)
considered in the previous subsection by the corresponding interacting fields,
which are formal power series of unbounded operators in the Fock space 
${\cal K}$ of free fields. Due to $[Q_u,{\cal L}(x)]=0$ (3.1), we can 
normalize the time ordered products such that
$${\bf (N6)}\quad\quad\quad\quad\quad\quad\quad\quad 
[Q_u,T({\cal L}(x_1)...{\cal L}(x_n))]=0\quad\quad\quad $$ 
$${\rm and}\quad\quad
[Q_u,T({\cal L}(x_1)...{\cal L}(x_n)F(x))]=\delta (F)T({\cal L}(x_1)...
{\cal L}(x_n)F(x)).$$
Hence $[Q_u,F_{{\rm int}\>{\cal L}}]=\delta (F)F_{{\rm int}\>{\cal L}}$ by 
(2.8-9). The fundamental normalization condition concerning the ghost number 
is (B.6) in combination with {\bf (N3)}; they imply {\bf (N6)}.
Again we fix the region ${\cal O}$ to be the double cone
${\cal O}\=d {\cal O}_{(-r,\vec 0),(r,\vec 0)},\>(r>0)$ (3.18) and assume 
the switching function $g\in {\cal D}({\bf R}^4)$ to be constant on ${\cal O}$
(3.19). We study the algebra $\tilde {\cal F}({\cal O})$ (3.17) of interacting
fields localized in ${\cal O}$. The ghost fields do not couple in QED, hence 
$$u_{{\rm int}\>{\cal L}}(x)=u(x),\quad\quad\quad
\tilde u_{{\rm int}\>{\cal L}}(x)=\tilde u(x).\eqno(5.9)$$

The abelian BRST-transformation $\tilde s=s_0+gs_1$ [BRS] should
be a graded derivation with zero 
square and compatible with the $*$-operation. In addition it 
shall induce the following transformations on the basic 
fields,\footnote{$^6$}{In contrast to the free case 
$\psi_{{\rm int}\>{\cal L}}$
and $\psq_{{\rm int}\>{\cal L}}$ 
are not observables in the sense of sect. 4.1. 
This different behaviour
can be understood physically by the accompanying soft photon cloud and
mathematically by Gauss' law.}
$$\tilde s(A^\mu_{{\rm int}\>{\cal L}}(x))=
i\d^\mu u(x),\quad\quad \tilde s(u(x))=0,
\quad\quad \tilde s(\tilde u(x))=-i\d_\mu A^\mu_{{\rm int}\>{\cal L}}(x),$$
$$\tilde s(\psi_{{\rm int}\>{\cal L}}(x))=
-g(x)\psi_{{\rm int}\>{\cal L}}(x)u(x),\quad\quad\tilde
s(\psq_{{\rm int}\>{\cal L}}(x))=
g(x)\psq_{{\rm int}\>{\cal L}}(x)u(x).\eqno(5.10)$$
(The pointwise products are well defined.)
Let us assume that we have constructed the interacting 
Kugo-Ojima charge $\tilde Q
=Q_{\rm int}(g)$. Then we shall define $\tilde s$ 
in terms of the corresponding current
such that $Q_{\rm int}(g)$ implements $\tilde s$ 
$$\tilde s(F)=Q_{\rm int}(g)F
-(-1)^{\delta (F)}FQ_{\rm int}(g),\quad\quad
F\in\tilde {\cal F}({\cal O}). \eqno(5.11)$$
If $Q_{\rm int}(g)$ is hermitian, $\tilde s$ is compatible 
with the $*$-operation, and ${\tilde s}^2=0$ is implied by 
$Q_{\rm int}(g)^2=0$. 

To get $Q_{\rm int}(g)$ we follow Kugo and Ojima [KO] 
and replace the current in the free charge 
$Q$ (5.5)  by the corresponding interacting
field $\d_\mu A^\mu_{{\rm int}\>{\cal L}}(x) {\dl}_\nu^x 
u(x)$. By means of the field equation (3.8) and the 
Ward identity (3.13) we find
$$\d^\nu_x[\d_\mu A^\mu_{{\rm int}\>{\cal L}}(x) {\dl}_\nu^x u(x)]=
-(\w\d_\mu A^\mu_{{\rm int}\>{\cal L}}(x))u(x)=(\d_\mu g)(x)
j^\mu_{{\rm int}\>{\cal L}}(x)u(x).\eqno(5.12)$$
Hence the current is conserved in the region where $g$ is 
constant. We may therefore define $\tilde s$ on an 
algebra $\tilde {\cal F}(\cal O)$ in the following way: we choose $g(x)=e=
{\rm const}$ on a neighbourhood ${\cal U}$ of $\bar {\cal O}$ and set
$${\tilde s}(F)\=d\int_{x_0=0} d^3x\,[\d_\mu A^\mu_{{\rm int}\>{\cal 
L}}(x) {\dl}_0 u(x),F]_\mp,\quad\quad
F\in\tilde {\cal F}({\cal O}).\eqno(5.13)$$
Because of current conservation, $\tilde s$ is implemented 
by the operators
$$Q_{\rm int}(g,k)=\int d^4x\,k^{\mu}(x)(\d_{\nu}A^{\nu}_{{\rm 
int}\>{\cal L}}(x)){\dl}_{\mu}^xu(x)\eqno(5.14)$$ 
with $k^{\mu}\in{\cal D}({\cal U})$ where 
$k^{\mu}-\delta^{\mu}_0 h=\d^{\mu}f$ for some 
$f\in{\cal D}({\cal U})$ and where $h \in{\cal D}({\cal U})$ is 
a suitably chosen smeared characteristic function
of the surface $\{ (0,{\bf x}),|{\bf x}|\le r\}$.

Now we are well prepared to prove that the definition (5.13) of ${\tilde s}$
agrees with the usual expressions (5.10) on the basic fields,
and to compute all further
commutators of $Q_{\rm int}(g)$ with the interacting sub Wick monomials
of ${\cal L}^0$ (3.15).

{\bf Theorem 7}: We assume that the interacting fields are normalized as
described in sections 2 and 3, especially that they
fulfil the field equations (3.8-9) and the Ward identities {\bf (N5)}. 
Furthermore we assume $g=e={\rm const}$ on  
the double cone 
${\cal O}={\cal O}_{(-r,\vec 0),(r,\vec 0)}$. Let $k^{\mu}$ 
as before.  Then we find the commutation rules
$$[Q_{\rm int}(g,k),A^\mu_{{\rm int}\>{\cal L}}(y)]=i\d^\mu u(y),\quad\quad
\quad [Q_{\rm int}(g,k),\d_\mu A^\mu_{{\rm int}\>{\cal L}}(y)]=0,
\eqno(5.15a,b)$$
$$[Q_{\rm int}(g,k),\psi_{{\rm int}\>{\cal L}}(y)]=
-e\psi_{{\rm int}\>{\cal L}}(y)u(y),
\quad\quad [Q_{\rm int}(g,k),\psq_{{\rm int}\>{\cal L}}(y)]=
e\psq_{{\rm int}\>{\cal L}}(y)u(y),\eqno(5.16a,b)$$
$$\{Q_{\rm int}(g,k),u(y)\}=0,\quad\quad\quad \{Q_{\rm int}(g,k),\tilde u(y)\}
=-i\d_\mu A^\mu_{{\rm int}\>{\cal L}}(y),\eqno(5.17a,b)$$
$$[Q_{\rm int}(g,k),F^{\mu\nu}_{{\rm int}\>{\cal L}}(y)]=0,\quad\quad\quad
[Q_{\rm int}(g,k),j^\mu_{{\rm int}\>{\cal L}}(y)]=0,\eqno(5.18a,b)$$
$$[Q_{\rm int}(g,k),(\gamma_\mu A^\mu\psi)_{{\rm int}\>{\cal L}}(y)]=-e
(\gamma_\mu A^\mu\psi)_{{\rm int}\>{\cal L}}(y)u(y)+
i\gamma_\mu\psi_{{\rm int}\>{\cal L}}(y)\d^\mu u(y),\eqno(5.19)$$
$$[Q_{\rm int}(g,k),(\psq\gamma_\mu A^\mu)_{{\rm int}\>{\cal L}}(y)]=e
(\psq\gamma_\mu A^\mu)_{{\rm int}\>{\cal L}}(y)u(y)+
i\psq_{{\rm int}\>{\cal L}}(y)\gamma_\mu
\d^\mu u(y),\eqno(5.20)$$
$$[Q_{\rm int}(g,k),(:\psq\gamma_\mu\psi:A^\mu)_{{\rm int}\>{\cal L}}(y)]=
(:\psq\gamma_\mu\psi:)_{{\rm int}\>{\cal L}}(y)i\d^\mu u(y),\eqno(5.21)$$
where always $y\in {\cal O}$.

{\it Proof}:
Since the ghost fields are not influenced by the interaction, we know that the 
ghost and antighost fields commute with all other 
interacting fields. Moreover, the pointwise products of these 
fields with a ghost or antighost field are well defined and 
behave in commutators as ordinary products of operators 
in spite of their character as operator valued 
distributions. (This may be verified by using techniques 
from microlocal analysis as exposed in [BF].) Thus the above 
relations follow from the commutation rules with $\d_{\nu}A^{\nu}_{{\rm 
int}\>{\cal L}}(x)$ in Proposition 3 and the ghost 
antighost anticommutation relations in equation (3.2). 
With these preparations the commutators (5.15-16) can easily be computed,
for example
$$[Q_{\rm int}(g),\psi_{{\rm int}\>{\cal L}}(y)]=
\int_{x_0=0} d^3x\,[\d_\mu 
A^\mu_{{\rm int}\>{\cal L}}(x),\psi_{{\rm int}\>{\cal L}}(y)]{\dl}_0^x u(x)=$$
$$=e\psi_{{\rm int}\>{\cal L}}(y)\int d^3x\,
D(x-y){\dl}_0^x u(x)=-e\psi_{{\rm int}\>{\cal L}}(y)u(y),\eqno(5.22)$$
where we have inserted Proposition 3, (3) and (A.31). By using other
commutators of $\d_\mu A^\mu_{{\rm int}\>{\cal L}}$ (Proposition 3)
we analogously prove (5.15a,b), (5.16b) and (5.21).
Alternatively (5.15b) can be obtained by taking the divergence
$\d_\mu^y$ of (5.15a). Part (a) of (5.17) is obvious due to $\{u(x),u(y)\}=0$;
let us compute part (b)
$$\{Q_{\rm int}(g),\tilde u(y)\}=-i\int_{x_0=0} d^3x\,
\d_\mu A^\mu_{{\rm int}\>{\cal L}}(x){\dl}_0^x D(x-y).\eqno(5.23)$$
From (5.12) we know
$\w \d_\mu A^\mu_{{\rm int}\>{\cal L}}(x)=0,\>\forall x\in {\cal O}$.
Therefore, we may apply (A.31) to (5.23), which yields (5.17b).
By applying $\d^\nu_y$ to (5.15a) and using $F^{\mu\nu}_{{\rm int}\>{\cal L}}=
\d^\mu A^\nu_{{\rm int}\>{\cal L}}-\d^\nu 
A^\mu_{{\rm int}\>{\cal L}}$ we get (5.18a).
Analogously by working with the field equations for 
$A^\mu_{{\rm int}\>{\cal L}}$
(3.8) and $\psi_{{\rm int}\>{\cal L}}$ (3.9),
we obtain (5.18b) and (5.19-20) from (5.15a) and (5.16a,b). $\quad\quad\w$

In the formal adiabatic limit, $\d_{\nu}A^{\nu}_{{\rm 
int}\>{\cal L}}$ converges to  $\d_{\nu}A^{\nu}$  (3.14) and 
therefore one expects that $Q_{\rm int}(g)$ will converge 
to the free Kugo-Ojima charge $Q$. Whereas in QED this 
reasoning seems to be correct, the corresponding argument does 
not work in nonabelian gauge theories (as can be seen by
an explicit calculation of the first order of $Q_{\rm int}(g)$) [BDF].

We therefore prefer not to work in the adiabatic limit. The 
price to pay is that $Q_{\rm int}(g)$ does  not agree with $Q$, 
so for the construction of the physical Hilbert space we 
have to check the conditions of Section 4. We easily find 
that $Q_{\rm int}(g,k)$ is hermitian for real valued $k$ 
and we even get
the nilpotency of $Q_{\rm int}(g,k)$,
$$Q_{\rm int}(g,k)^2={1\over 2}\{Q_{\rm int}(g,k),Q_{\rm int}(g,k)\}=$$
$$={1\over 2}\int d^4x\,h(x)\int d^4y\,h(y)
[\d_\mu A^\mu_{{\rm int}\>{\cal L}}(x),\d_\nu A^\nu_{{\rm int}\>{\cal L}}(y)]
{\dl}_0^x {\dl}_0^y u(x)u(y)=0,\eqno(5.24)$$
by means of $k^\mu=\delta^\mu_0 h+\d^\mu f$ and
Proposition 3, (2).\footnote{$^7$}{We recall that the commutator
$[\d A_{{\rm int}\>{\cal L}}(x),\d A_{{\rm int}\>{\cal L}}(y)]$ vanishes
for {\it all} $x$ and $y$ for which $\supp \partial_{\mu}g$ does not
intersect $\overline{{\cal O}_{x,y}\cup {\cal O}_{y,x}}$.} 

But we need in addition that the zeroth order term $Q_0(k)$ of $Q_{\rm 
int}(g,k)$ (5.14) satisfies the positivity assumption 
(4.8). There seems to be no reason why this should 
hold for a generic choice of $k$. One might try to control 
the limit when $k$ tends to a smeared characteristic 
function of the $t=0$ hyperplane (in order that $Q_0(k)$ becomes equal to
the free charge Q (5.5)), but without a priori information on the existence
of an $\tilde s$-invariant state this appears to be a hard problem.    
 
There is a more elegant way to get rid of these problems 
which relies on the local character of our construction. We 
may embed our double cone  ${\cal O}$ isometrically into 
the cylinder  ${\bf R}\times C_L$, where $C_L$ is a cube of 
length $L,\quad L\gg r$, with suitable boundary 
conditions (see appendix A), and where the first factor denotes the time 
axis.  If we choose the compactification length $L$ 
big enough, the physical
properties of the local algebra $\tilde {\cal F}({\cal O})$ are not changed.

The quantization of the free fields on this cube is worked
out in appendix A.
The inductive construction of the perturbation series for the S-matrix or 
the interacting fields is not changed by the 
compactification, sections 2 and 3 
can be adopted without any modification [BF]. We assume the switching function 
$g$ to fulfil
$$g(x)=e={\rm constant}\quad\quad\forall x\in {\cal O}\cup
\{(x_0,\vec x)|\,|x_0|<\epsilon\}\quad\quad (r\gg\epsilon>0)\eqno(5.25)$$
on ${\bf R}\times C_L$ and to have compact support in 
timelike directions. Now we may insert 
$$k^{\mu}(x):=\delta^{\mu}_{0}h(x_0),
\quad\quad\quad h\in {\cal D}([-\epsilon,\epsilon]),
\quad\quad\quad\int dx_0\,h(x_0)=1$$
into the expression
(5.14) for $Q_{\rm int}$, because $(x_0,\vec x)\rightarrow h(x_0)$ is 
an admissible test function on ${\bf R}\times C_L$. We define
$$Q_{\rm int}(g):{\cal D}\longrightarrow {\cal D},\quad\quad
Q_{\rm int}(g)\=d \int dx_0\,h(x_0)\int_{C_L} d^3x\,
\d_\mu A^\mu_{{\rm int}\>{\cal L}}(x){\dl}_0^x u(x).\eqno(5.26)$$
By means of (5.12) and the fact that $g$ 
is constant on the region of integration
(the time-slice $[-\epsilon,\epsilon]\times C_L$), we conclude that
the result of the integration
over $C_L$ is independent of $x_0$ and hence the arbitrariness
in the choice of $h(x_0)$ drops out,
$$Q_{\rm int}(g)=\int_{C_L,x_0={\rm const.},|x_0|<\epsilon} d^3x\,
\d_\mu A^\mu_{{\rm int}\>{\cal L}}(x){\dl}_0^x u(x).\eqno(5.27)$$

By construction, $Q_{\rm int}(g)$ implements the BRST-transformation
(5.13) and fulfills (see (5.24))
$$Q_{\rm int}(g)^2=0,\quad\quad\quad Q_0=Q,\quad\quad\quad [Q_u,Q_{\rm int}(g)]
=-Q_{\rm int}(g)\quad {\rm and}\quad Q_{\rm int}(g)^*=
Q_{\rm int}(g),\eqno(5.28)$$
where $Q_0$ is the zeroth order and $Q$ is the free charge (5.5).
The last property relies on the *-selfadjointness of $u$ and
$A^\mu_{{\rm int}\>{\cal L}}$.

We emphasize that our construction describes locally QED also in the {\it
non-compactified} Minkowski space (this is the main concern of the paper)
and, therefore, should not depend on the compactification
length $L$. On the level of the algebras this is evident. The local algebras
of interacting fields or observables belonging to different values of $L$
are {\it isomorphic}. We conjecture that also the state space
(i.e. the set of expectation functionals induced by vectors in the
physical Hilbert space) is independent of $L$, but this remains to be proven.
\vskip 1cm
{\trm 6. Outlook}
\vskip 1cm
In an abstract setting,
Buchholz has developed concepts for the treatment of scattering of
infraparticles [Bu1]. It would be interesting to apply his ideas to
perturbative QED. Our construction may be helpful for such an
investigation which might lead to a more satisfactory understanding of
the infrared problem in QED.

The importance of a {\it local} construction of the 
observables becomes even more evident in nonabelian gauge
theories. There, in the absence of Higgs fields, the 
adiabatic limit seems not to exist and, hence, only a {\it local} 
model makes sense in the framework of perturbation theory. In the
application of
our construction to nonabelian gauge theories some technical problems appear,
but we see no principal obstacle [BDF].

This is also the perspective for the generalization to curved space-times, 
where the techniques of [BF] can be used.

An open question is the physical meaning of the remaining 
normalization conditions in a local perturbative 
construction, after the restrictions from gauge invariance and other
symmetries are taken into account.
The parameters involved may be considered as 
structure constants of the algebra of
observables, but their usual interpretation as charge and mass
involve the adiabatic limit.
\vfill
\eject
{\trm Appendix A: Implementation of the free 
BRST-transformation on the spatially
compactified\break Minkowski space}
\vskip 1cm
In this appendix we quantize the free gauge, 
ghost and spinor fields in a finite 
spatial volume. Special care is needed in the choice of boundary conditions.
In the second part of this appendix we prove that the free Kugo-Ojima charge
$Q$ (5.5) is nilpotent, implements the BRST-transformation of free fields 
and fufills our positivity assumption (4.8).

Let $C_L$ be the open cube of length $L$. The algebra of a free scalar 
field $\varphi_L$ with mass $m\geq 0$ on 
${\bf R}\times C_L$ with {\it Dirichlet 
boundary conditions} \footnote{$^8$}{We first tried {\it periodic} boundary 
conditions, but this seems not to work for massless particles because of the 
existence of zero modes [DF]. For bosonic particles the 
algebra of the zero mode 
agrees with the algebra of a free Schr\"odinger particle in one dimension. 
There is no ground state on this algebra, 
and this makes it impossible to define 
the physical Hilbert space as the cohomology of the free Kugo-Ojima charge $Q$.
Therefore, we choose boundary conditions which exclude the zero mode.}
is the unital *-algebra generated by elements 
$\varphi_L(f)$, $f\in {\cal D}({\bf R}\times C_L)$ with the 
relations 
$$f\mapsto\varphi_L(f) {\rm \ is \ linear }\eqno(A.1)$$
$$\varphi_L((\w +m^2)f)=0\eqno(A.2)$$
$$\varphi_L(f)^{*}=\varphi_L(\bar f)\eqno(A.3)$$
$$[\varphi_L(f),\varphi_L(g)]=-i\int 
d^4x\,d^4y\,f(x)g(y)\Delta_{L}(x,y)\eqno(A.4)$$
where $\Delta_{L}$ is the fundamental solution of the Klein 
Gordon equation on ${\bf R}\times C_L$ with Dirichlet boundary 
conditions, which has the explicit form
$$\Delta_{L}(x^0,{\bf x},y^0,{\bf y)}=\sum_{s\in 
S}(-1)^{n(s)}\Delta(x^0-y^0,{\bf x}-s({\bf y}))\eqno(A.5)$$
where $S$ is the group generated by the reflections on the 
planes which bound $C_L$ and $n(s)$ is the number of 
reflections occuring in $s$ (which is well defined modulo 2). 
In particular one sees that $\Delta_{L}$ coincides with 
$\Delta$ on ${\cal O}\times {\cal O}$ if the closure of the double cone 
${\cal O}$ is contained in ${\bf R}\times C_L$, considered 
as a region in Minkowski space. {\it Hence the algebra ${\cal 
F}({\cal O})$ associated 
to ${\cal O}$ is independent of the boundary conditions.}

Since $\Delta_{L}$ depends only on time differences, the 
algebra is invariant under time translations. The ground state 
is the quasifree state whose 2-point function is the 
positive frequency part $\Delta_{L}^+$ of $\Delta_{L}$. In the massive 
case it is given in terms of the Minkowski space 2-point 
functions by a formula analogous to the formula for the 
commutator function. In the massless case a corresponding 
sum might not converge, instead we exploit the fact that the 
possible frequencies must have squares which are eigenvalues 
of the Laplace operator on $C_L$ with Dirichlet boundary 
conditions. In particular zero modes do not appear. 
Therefore the frequency splitting can be done by 
convolution with the distribution 
$$H(t)={ih(t)\over (2\pi)^{1\over 2}(t+i\varepsilon)}\eqno(A.6)$$ 
where $h$ is a test function 
from the Schwartz space ${\cal S}({\bf R})$ with $h(0)=1$ 
and a Fourier transform with support in the interval 
$(-\omega,\omega)$ where $\omega^2$ is the smallest 
eigenvalue of the Laplace operator.
$H*\Delta$ differs from $\Delta^{+}$ by a smooth function 
and decays fast in spatial directions. Hence $\Delta_{L}^+$ 
admits a representation in terms of $H*\Delta$,
$$\Delta_{L}^+(x^0,{\bf x},y^0,{\bf y)}=\sum_{s\in 
S}(-1)^{n(s)}H*\Delta(x^0-y^0,{\bf x}-s({\bf y})),\eqno(A.7)$$   
and $\Delta_{+}-\Delta_{L}^+$ is smooth on ${\cal O}\times {\cal O}$.   
Due to a result of Verch [V], this implies that the representation of  
${\cal F}({\cal O})$ induced by the ground state 
on ${\bf R}\times C_L$ is unitarily 
equivalent to the Minkowski space vacuum representation. 
 
We also need to compare the Wick products in both 
representations. Since the two point functions differ by 
a smooth function we may use the Minkowski space definition 
of Wick products also on ${\bf R}\times C_L$. \footnote{$^{9}$}{
To see this we consider
$$:\phi (x_1)...\phi (x_n):_\omega\Psi= {\delta^n\over i^n\delta f(x_1)...
\delta f(x_n)}e^{{1\over 2}\omega (f,f)}e^{i\phi (f)}\Psi \vert_{f=0}=$$
$$=\sum_{l=0}^{[{n\over 2}]}\sum {\delta^{2l} e^{{1\over 2}[\omega (f,f)
-\omega^\prime (f,f)]}\over i^{2l}\delta f(x_{i_1})
\delta f(x_{j_1})...\delta f(x_{i_l})
\delta f(x_{j_l})}\vert_{f=0}\cdot {\delta^{(n-2l)}e^{{1\over 2}\omega^\prime 
(f,f)}e^{i\phi (f)}\Psi\over i^{n-2l}\delta f(x_1)...\hat i_1\hat j_1...
\hat i_l\hat j_l...\delta f(x_n)}\vert_{f=0}=$$
$$=:\phi (x_1)...\phi (x_n):_{\omega^\prime}\Psi+
\sum_{l=1}^{[{n\over 2}]}(-1)^l
\sum (\omega-\omega^\prime)(x_{i_1}-x_{j_1})...(\omega-\omega^\prime)
(x_{i_l}-x_{j_l})\cdot$$
$$\cdot :\phi (x_1)...\hat i_1\hat j_1...\hat i_l\hat j_l...
\phi (x_n):_{\omega^\prime}\Psi,$$
where $\omega$ and $\omega^\prime$ denote quasifree 
states or the corresponding 
two-point functions, $\Psi$ is a suitable state vector 
and the hat means the omission
of the corresponding factor. Hence  if 
$(\omega-\omega^\prime)$ is a smooth function, 
the limit $(x_i-x_j)\rightarrow 0\quad\forall 
i\not=j$ exists in $:\phi (x_1)...
\phi (x_m):_\omega\quad \forall m\in {\bf N}$ iff it exists in $:\phi (x_1)...
\phi (x_m):_{\omega^\prime}\quad \forall m\in {\bf N}$.} In [BF]  
a domain of definition of Wick polynomials was found which 
depends only on the equivalence class of the representation, 
hence also the Wick products can be identified. 

For the electromagnetic field we may use {\it metallic boundary 
conditions}, i.e. the pullback of the 2-form $F$ vanishes at 
the boundary (which means that the tangential components of 
the electric field 
and the normal component of the magnetic field vanish). In 
addition we assume that the auxiliary Nakanishi-Lautrup 
field $B=\d^{\mu}A_{L\,\mu}$ (in Feynman gauge) satisfies 
Dirichlet boundary conditions.

The corresponding commutator function is
$$D_{\mu\nu,L}(x^0,{\bf x},y^0,{\bf y)}=\sum_{s\in 
S}(-1)^{n(s)}s_{\mu}^{\lambda}g_{\lambda\nu}
D(x^0-y^0,{\bf x}-s({\bf y}))\eqno(A.8)$$
where the matrix $(s_{\mu}^{\lambda})$ describes the action 
of $s$ on covectors, e.g. for a reflection $s$ on a plane $x_2={\rm const.}$
we have $s_{\mu}^{\lambda}g_{\lambda\nu}={\rm diag}(1,-1,1,-1)_{\mu\nu}$.
The algebra generated by the vector 
potential $A_{L\,\mu}$ can then be defined as in the scalar 
case, and again the subalgebra associated to the double 
cone ${\cal O}$ is independent of the boundary conditions. A 
ground state can also be defined in terms of the positive 
frequency part of the two point function; as on Minkowski 
space, it violates the positivity condition.

The ghost and antighost fields are quantized with Dirichlet boundary
conditions, i.e. by the relation
$$(u_L(f)+i\tilde u_L(g))^2=\int d^4x\,d^4y\,f(x)g(y)D_{L}(x,y)\eqno(A.9)$$
which replaces the commutator condition. 
($D_L$ is obtained from $\Delta_L$ (A.5)
by setting $m=0$.)
The ground state is 
obtained from the two point function
$$\omega_{u}(\tilde u_L(x)u_L(y)=iD_{L}^{+}(x,y)\eqno(A.10)$$
which again violates positivity w.r.t. the *-operation, 
which is defined by (cf. (3.5))
$$(u_L(f)+i\tilde u_L(g))^*\=d(u_L(\bar f)+i\tilde u_L(\bar g)).\eqno(A.11)$$

Finally, we have to find suitable boundary conditions for 
the electron field. For simplicity, we 
choose {\it periodic boundary conditions}.
Because they are invariant under charge conjugation, the expectation value of 
the electric current (normal ordered w.r.t. the 
Minkowski vacuum) vanishes in the
ground state (of the cube), hence the 
interaction Lagrangian ${\cal L}^0$ (3.15)
keeps the same form as on Minkowski space.

We are now going to represent these algebras in Fock spaces. The one-particle
Hilbert space ${\cal H}$ for a massless free scalar field is the completion
with respect to the scalar product
$$(f,g)=i\int_{C_L,x_0={\rm const.}}d^3x\,f(x)^*{\dl}_0^x g(x)\eqno(A.12)$$
of the space 
of all smooth functions $f:{\bf R}\times C_L\rightarrow {\bf C}$ which are 
positive frequency solutions of the wave equation 
and fulfil the considered boundary conditions.
For Dirichlet boundary conditions the mode decomposition for the functions 
$f\in {\cal H}$ reads
$$f(x)=\sum_{n_1,n_2,n_3=1}^\infty f_{\vec n}v_{\vec n}(x),
\quad\quad f_{\vec n}={\rm const.},\eqno(A.13)$$
where 
$$v_{\vec n}(x)\=d {2\over (\omega_{\vec n}L^3)^{1\over 2}}\,
{\rm sin}\,k_{n_1}x_1\>\,
{\rm sin}\,k_{n_2}x_2\>\,{\rm sin}\,k_{n_3}x_3\>\,
e^{-i\omega_{\vec n}x_0},\quad 
{\vec k}_{\vec n}\=d {\pi\over L}
{\vec n},\quad\omega_{\vec n}\=d \Vert {\vec k}_{\vec n}\Vert 
={\pi\over L}\Vert {\vec n}\Vert \eqno(A.14)$$
(the normalization is such that $(v_{\vec n},v_{\vec m})=\delta_{{\vec n},
{\vec m}}$).

The representation of gauge fields in
Feyman gauge requires an indefinite inner product space. 
We describe it in terms of a Krein operator
$$J_L=(-1)^{N_0}\otimes 1\otimes J_g\quad\quad {\rm on}\quad {\rm the}
\quad {\rm Hilbert}\quad {\rm space}
\quad\quad {\cal K}_L={\cal K_A}\otimes {\cal K_\psi}\otimes {\cal K}_g. 
\eqno(A.15)$$
The latter is the tensor product of the photon-, spinor- and ghost-Fock space.
$N_0$ is the particle number operator for scalar photons and $J_g$ will be 
defined below. The Krein operator (A.15) fulfils
$$J_L^2=1,\quad\quad\quad\quad J_L^+=J_L\eqno(A.16)$$
($^+$ denotes the adjoint in ${\cal K}_L$),
and the dense invariant domain ${\cal D}_L$ 
can be chosen such that $J_L{\cal D}_L
={\cal D}_L$. The indefinite inner product is given by
$$<a,b>\=d (a,J_Lb),\quad\quad a,b\in {\cal K}_L,\eqno(A.17)$$
where $(.,.)$ denotes the (positive definite) scalar product in ${\cal
K}_L$, and
the *-operation with respect to (A.17) is
$$O^*\=d J_LO^+J_L,\quad\quad <Oa,b>=<a,O^*b>.\eqno(A.18)$$
Let $a_{\vec n}^\mu, a_{\vec n}^{\mu +}, c_{j\vec n}, c_{j\vec n}^+\>(j=1,2)$ 
be the usual annihilation and creation operators of the 
Fock spaces ${\cal K_A}$
and ${\cal K}_g$ which fulfil the (anti-)commutation relations
$$[a_{\vec n}^\mu,a_{\vec m}^{\nu +}]=\delta_{{\vec n},{\vec m}}
\delta^{\mu\nu}L^3 2\omega_{\vec n}\eqno(A.19)$$
and
$$\{c_{j\vec n},c_{l\vec m}^+\}=\delta_{{\vec n},{\vec m}}
\delta_{jl}L^3 2\omega_{\vec n}.\eqno(A.20)$$
The ghost fields $u_L$ and $\tilde u_L$ and the zeroth component of 
the photon field $A_L^0$ are scalar fields with 
Dirichlet boundary conditions and 
some unusual sign conventions 
$$\tilde u_L(x)=\sum_{n_1,n_2,n_3=1}^\infty {1\over
(2\omega_{\vec n}L^3)^{1\over 2}}(-c_{1\vec n}v_{\vec n}(x)+
c_{2\vec n}^{+}v_{\vec n}(x)^*),\eqno(A.21)$$
$$u_L(x)=\sum_{n_1,n_2,n_3=1}^\infty {1\over
(2\omega_{\vec n}L^3)^{1\over 2}}(c_{2\vec n}v_{\vec n}(x)+
c_{1\vec n}^{+}v_{\vec n}(x)^*),\eqno(A.22)$$
$$A_L^0(x)=\sum_{n_1,n_2,n_3=1}^\infty {1\over 
(2\omega_{\vec n}L^3)^{1\over 2}}(a_{\vec n}^0v_{\vec n}(x)-
a_{\vec n}^{0+}v_{\vec n}(x)^*).\eqno(A.23)$$
The normalizations are such that they go 
over into the usual Lorentz covariant conventions of the non-compactified space
by replacing $({2\pi\over L})^3\sum_{\vec n\in {\bf Z}^3,\vec n\not= \vec 0}$
by $\int d^3k$. For the spatial components of the photon field $A_L^\mu$
we have a mixture of Dirichlet and von Neumann boundary conditions. For
example for $\mu=2$ we define
$$v_{2\vec n}(x)\=d {\eta_{\vec n}2\over (\omega_{\vec n}L^3)^{1\over 2}}
{\rm sin}\,k_1x_1\>\,{\rm cos}\,k_2x_2\>\,
{\rm sin}\,k_3x_3\>\,e^{-i\omega_{\vec n}x_0},
\quad n_1,n_3=1,2,...,\quad n_2=0,1,2,...,\eqno(A.24)$$
where $\eta_{\vec n}=1$ for $n_2\geq 1$ and 
$\eta_{\vec n}=2^{-{1\over 2}}$ for 
$n_2=0$ and similar for $\mu=1,3$. Then we set
$$A_L^l(x)=\sum_{\vec n}{1\over
(2\omega_{\vec n}L^3)^{1\over 2}}(a_{\vec n}^lv_{l\vec n}(x)+
a_{\vec n}^{l+}v_{l\vec n}(x)^*),\quad\quad l=1,2,3.\eqno(A.25)$$
$J_g$ (A.15) is defined implicitly by (A.11), (A.16) and 
(A.21-22), i.e. we have
$$c_{1\vec n}^*=c_{2\vec n}^{+},\quad\quad c_{2\vec n}^*=c_{1\vec n}^{+}.$$
For the photon two-point function we obtain
$$(\Omega_A,A^{\mu\,+}_L(x)A^\nu_L(y)\Omega_A)=
<\Omega_A,A^\mu_L(x)J_L A^\nu_L(y)\Omega_A>=
\delta^{\mu\nu}\sum_{\vec n}
v_{\mu\vec n}(x)v_{\mu\vec n}(y)^*,\quad\> 
(v_{0\vec n}\=d v_{\vec n})\eqno(A.26)$$
which is obviously positive.

We now transfer the construction of the interacting 
fields (sections 2 and 3) from
Minkowski space to ${\bf R}\times C_L$. Due to [BF] 
there is no principle obstacle
and there are only few changes in the formulas. Since 
spatial translation invariance
is lost, the commutator functions and propagators do not only depend on the 
relative coordinates, they must be replaced by the 
above given expressions (A.5), (A.7)
(A.8) etc.. Some care is required in the proof of Proposition 3. To get
(3.22) we use the identity
$$\d_\mu^x D_L^{{\rm av}\,\mu\rho}(x,z_m)=-\d_{z_m}^\rho D_L^{\rm av}(x,z_m)
\eqno(A.27)$$
(which is an immediate consequence of (A.5), 
(A.8)) and the fact that the boundary 
terms of the 'partial integration' vanish because $D_L^{\rm av}$ (A.5) fulfills
Dirichlet boundary conditions. By Lemma 1 {\bf (A)} we obtain
$$A_{{\rm int}\>{\cal L}}^\mu (x)=A^\mu (x)
+\sum_{n=1}^{\infty}{i^{n+1}\over n!}\int d^4x_1...d^4x_n\,\cdot$$
$$\cdot\sum_{l=1}^ng(x_1)...g(x_n)D^{\mu\nu\,{\rm ret}}_L(x,x_l)
R({\cal L}^0(x_1)...\hat l...{\cal L}^0(x_n);j^\mu (x_l)).\eqno(A.28)$$
Since $D^{\mu\nu\,{\rm ret}}_L(x,x_l)$ fulfills the boundary conditions 
of $A^\mu(x)$ we conclude that $A_{{\rm int}\>{\cal L}}^\mu$ {\it obeys
the same boundary conditions as the corresponding free field}, and similar
for $\psi_{{\rm int}\>{\cal L}},\,u_{{\rm int}\>{\cal L}}$, etc..

Let us turn to the implementation of the free BRST-transformation.
In the following and in the main text we omit the lower index 'L'.
Due to $\d^\mu [(\d_{\nu}A^{\nu}){\dl}_\mu u]=0$ the definition
$$Q\=d \int_{C_L,\,x_0={\rm const.}} d^3x\,
(\d_{\nu}A^{\nu}(x)){\dl}_0u(x)\eqno(A.29)$$
of the free Kugo-Ojima charge (5.5) is independent of $x_0$.
Because of $A^{\mu\,*}=A^\mu,\,u^*=u$ (3.5) we immediately see $Q^*=Q$.
By means of
$$\int_{C_L,\,x_0={\rm const.}} d^3x\>D(y-x){\dl}_0^x\phi (x)=\phi (y),
\quad\quad\forall\phi\quad\quad {\rm fulfilling}\quad\quad\w\phi=0,
\eqno(A.30)$$
one proves that the charge $Q$ implements the BRST-transformation (5.2)
of the free fields, e.g.
$$[Q,A^\mu (y)]=\int_{C_L,\,x_0={\rm const.}} d^3x\,[\d_{\nu}A^{\nu}(x),
A^\mu (y)]{\dl}_0^x u(x)=$$
$$=-i\d_y^\mu \int_{C_L,\,x_0={\rm const.}} d^3x\,D(x-y){\dl}_0^x u(x)
=i\d^\mu u(y).\eqno(A.31)$$
The transformation (5.3-4) of Wick monomials and derivated fields is also
implemented by $Q$, because of
$$[Q,:\phi_1(x)\phi_2(x)...:]_\mp =:[Q,\phi_1(x)]_\mp\phi_2(x)...:+
(-1)^{\delta (\phi_1)}:\phi_1(x)[Q,\phi_2(x)]_\mp ...:+...\eqno(A.32)$$
and
$$[Q,\d_\mu\phi(x)]_\mp =\d_\mu^x [Q,\phi (x)]_\mp .$$
We easily find that $Q$ is nilpotent
$$Q^2={1\over 2}\{Q,Q\}=\int_{C_L,\,x_0={\rm const.}} d^3x\>
\{Q,(\d_{\nu}A^{\nu}){\dl}_0u\}=0.\eqno(A.33)$$

Inserting (A.22-23) and (A.25) into (A.29) we obtain
$$Q={1\over L^3 2^{1\over 2}}\sum_{n_1,n_2,n_3=1}^\infty
[c_{1\vec n}^+b_{1\vec n}+b_{2\vec n}^+c_{2\vec n}]\eqno(A.34)$$
in a straightforward way (the sum in $Q$ converges in the topology of the Krein
space on the dense invariant domain ${\cal D}$), where
$$b_{1\vec n}\=d {1\over 2^{1\over 2}}(a_{\vec n}^0+i{k_{\vec n}^j a_{\vec n}^j
\over \omega_{\vec n}}),\quad\quad b_{2\vec n}\=d {-1\over 2^{1\over 2}}
(a_{\vec n}^0-i{k_{\vec n}^j a_{\vec n}^j\over \omega_{\vec n}}),\eqno(A.35)$$
which implies
$$[b_{j\vec n},b_{l\vec m}^+]=\delta_{{\vec n},{\vec m}}
\delta_{jl}L^3 2\omega_{\vec n}\eqno(A.36)$$
(cf. sect. 5 of [DHS1]). By means of $Q^2=0$ one finds 
similarly to [K] that the 
dense invariant domain ${\cal D}$ has the decomposition
$${\cal D}={\rm Ra}\>Q\oplus ({\rm Ke}\>Q\cap {\rm Ke}\>Q^+)
\oplus {\rm Ra}\>Q^+
\eqno(A.37)$$
and these three subspaces are pairwise orthogonal with 
respect to the scalar product (.,.)
(cf. (A.17-18)). Additionally one easily verifies
$${\rm Ke}\>Q\cap {\rm Ke}\>Q^+={\rm Ke}\>\{Q,Q^+\}.\eqno(A.38)$$
Inserting (A.34) we find
$$\{Q,Q^+\}={1\over L^3}\sum_{n_1,n_2,n_3=1}^\infty\omega_{\vec n}
(b_{1\vec n}^+b_{1\vec n}+b_{2\vec n}^+b_{2\vec n}
+c_{1\vec n}^+c_{1\vec n}+c_{2\vec n}^+c_{2\vec n}),\eqno(A.39)$$
which agrees up to factors $2\omega_{\vec n}^2$ with the 
particle number operator of 
the ghosts and the longitudinal and scalar photons; 
however, the kernels completely agree.
Obviously the Krein operator $J$ (A.15) is the identity 
on ${\rm Ke}\>\{Q,Q^+\}$. 
Additionally $J$ maps ${\rm Ra}\>Q$ onto ${\rm Ra}\>Q^+$, due to $JQ=Q^+J$. 
From the decomposition
(A.37) we conclude that our positivity assumption (4.8) is 
in fact satisfied, i.e.
the indefinite product $<.,.>$ is positive 
semidefinite on ${\rm Ke}\>Q$ and the 
null vectors in ${\rm Ke}\>Q$ are precisely ${\rm Ra}\>Q$.

The vectors in ${\rm Ke}\>\{Q,Q^+\}$ are distinguished representatives of the 
equivalence classes in the physical 
space ${\cal H}={{\rm Ke}\>Q\over {\rm Ra}\>Q}$ (4.9).
They provide the usual physical picture,
namely that the states in ${\cal H}$ are built up from electrons, positrons and
transversal photons only. 
\vskip 1cm
{\trm Appendix B: Proof of the Ward identities}
\vskip 1cm
We recall the Ward identites (3.16)
$$\d_\mu^y T\Bigl( j^\mu (y)A_1(x_1)...A_n(x_n)\Bigr)=
i\sum_{j=1}^n\delta (y-x_j)
T\Bigl( A_1(x_1)...(\theta A_j)(x_j)...A_n(x_n)\Bigr),\eqno(B.1)$$
where 
$$(\theta A_j)\=d {d\over d\alpha}\vert_{\alpha =0} A_{j\alpha}=i(r_j-s_j)A_j
\quad\quad {\rm for}\quad\quad A_j=:\psi^{r_j}\psq^{s_j}B_1...B_l:\eqno(B.2)$$
($B_1,...,B_l$ are non-spinorial free fields, i.e. photon or ghost operators),
and $A_{j\alpha}$ is given by the global $U(1)$-transformation (3.17). Note
$$(\theta A_j)=-i[Q_\psi,A_j]\quad\quad {\rm with}\quad\quad
Q_\psi\=d\int d^3x\,:\psq(x)\gamma^0\psi(x):,\eqno(B.3)$$
i.e. $Q_\psi$ is the infinitesimal generator of the transformation (3.17).

There exist several proofs of the Ward identities in QED,
e.g. [FHRW,S]. Here we want to show that they can be fulfilled in our
framework; in particular we have to check that they are compatible
with our normalization conditions. We follow ideas from [St2].

First we point out a consequence of the Ward identities (B.1). For a given
$(x_1,...,x_n)$ let ${\cal O}\subset {\bf R}^{4}$
be a double cone which contains the points $x_1,\dots,x_n$ and let $g$
be a test function which is equal to 1 on a neighbourhood of
$\overline{\cal O}$. 
We decompose $\partial^{\mu} g=a^{\mu}-b^{\mu}$
such that $\supp a^{\mu}\cap (\overline{V}_-+{\cal O})=\emptyset$ and  
$\supp b^{\mu}\cap (\overline{V}_++{\cal O})=\emptyset$. We smear out (B.1)
with this $g$ in $y$. Then, by causal
factorization, the left hand side of (B.1) becomes
$$-j^{\mu}(a_{\mu})T\Bigl(A_1(x_1)...A_n(x_n)\Bigr)+
T\Bigl(A_1(x_1)...A_n(x_n)\Bigr)j^{\mu}(b_{\mu})=$$
$$=-[j^{\mu}(a_{\mu}),T\Bigl(A_1(x_1)...A_n(x_n)\Bigr)]-
T\Bigl(A_1(x_1)...A_n(x_n)\Bigr)j^{\mu}(\partial_{\mu}g).\eqno(B.4)$$
The second term vanishes because $j^{\mu}$ is a conserved current. Since
$T(A_1(x_1)...A_n(x_n))$ is localized in ${\cal O}$,
the term $-j^{\mu}(a_{\mu})$ in the commutator can be replaced by $Q_{\psi}$.
\footnote{$^{10}$}
{This may be seen as follows. Different choices of $a_{\mu}$ differ
only in the spacelike complement of ${\cal O}$ and therefore do not
affect the commutator. We may choose
$$a_{\mu}(x)=\partial_{\mu}g(x)\int_{-\infty}^{ex}h(t)dt \eqno(B.5)$$
where $e$ is a suitable timelike unit vector and $h$ is a test
function of one variable with sufficiently small support and total
integral 1, i.e. the integral in (B.5) is a ${\cal C}^\infty$-approximation
to $\Theta (ex-c)$ ($c\in {\bf R}$ is a suitable constant).
Then by current conservation and partial integration we obtain
$$-j^{\mu}(a_{\mu})=j^{\mu}(e_{\mu}gh(e\cdot))=\int dt
h(t)\int_{ex=t}g(x)
j^{\mu}(x)\varepsilon_{\mu\nu\rho\sigma}dx^{\nu}dx^{\rho}dx^{\sigma}, $$
hence the statement follows from $g\equiv 1$ on ${\cal O}$.}
Hence the validity of the following lemma is necessary for the Ward
identities:  

{\bf Lemma 8}: In agreement with {\bf (N1-4)} and {\bf (N6)}
the normalizations can be chosen such that the
vacuum expectation values of the time ordered products 
vanish, if the sum of the charges
of the factors is different from zero
$$<\Omega|T(A_1...A_n)|\Omega>=0\quad\quad\quad{\rm for}\quad\quad\quad
\sum_j(r_j-s_j)\not= 0.\eqno(B.6)$$
Under this condition the following identity becomes true
$$[Q_\psi,T\Bigl( A_1(x_1)...A_n(x_n)\Bigr)] =\sum_{j=1}^n
T\Bigl( A_1(x_1)...[Q_\psi,A_j(x_j)]...A_n(x_n)\Bigr)\equiv$$
$$\equiv
i\sum_{j=1}^n T\Bigl( A_1(x_1)...(\theta A_j)(x_j)...A_n(x_n)\Bigr).
\eqno(B.7)$$

{\it Proof of Lemma 8}: 
The lemma is certainly fulfilled for $n=1$, and we proceed
inductively with respect to the order $n$. For each fixed $n$ we consider
a second induction in the sum of the degrees of the Wick monomials $A_j,\>
j=1,...,n$. We commute the assertion (B.7) with the free fields. 
After inserting
{\bf (N3)} we can use the inductive assumption and find that 
these commutators vanish.
Therefore,  the identity (B.7) can only be violated by a C-number. (An
analogous computation is given below in step 1 of the proof of the 
Ward identities.)
To determine this C-number we consider the vacuum expectation value of (B.7). 
Since $Q_\psi=Q_\psi^*$ annihilates the vacuum we find
$<\Omega|[Q_\psi,T(A_1...A_n)]|\Omega>=0$. Moreover note
$$\sum_j <\Omega|T(A_1...(\theta A_j)...A_n)|\Omega>=
i\Bigl(\sum_j(r_j-s_j)\Bigr)
<\Omega|T(A_1...A_n)|\Omega>.\eqno(B.8)$$
Due to the causal factorization and the validity of (B.7) in lower orders,
the expression (B.8) must be local. Hence we can require (B.6) as a 
normalization condition,
i.e. we extend zero by zero to the total diagonal. Obviously this 
prescription is
compatible with {\bf (N1-4)} and {\bf (N6)}. 
This completes the proof of the lemma. $\quad\quad\w$

{\it Proof of the Ward identities}: We show that
all Ward identities can be satisfied by choosing a suitable normalization
of the vacuum expectation values
of the time ordered products which contain no free field factor
and with vanishing total charge (B.6).
We work with the same double inductive procedure as in the previous proof.

{\it Step 1:} Again we commute the assertion with the free fields.
By means of {\bf (N3)} we obtain
$$\Bigl[\{\d_\mu^y T\Bigl( j^\mu (y)A_1(x_1)...A_n(x_n)\Bigr)-
i\sum_m\delta (y-x_m)
T\Bigl( A_1(x_1)...(\theta A_m)(x_m)...A_n(x_n)\Bigr)\},\phi_j(z)\Bigr] =$$
$$=i\sum_k\Bigl\{\d_\mu^y T\Bigl( j^\mu (y)A_1...{\d A_k\over\d\phi_l}...
A_n\Bigr)-i\sum_{m\,(m\not= k)}\delta (y-x_m)
T\Bigl( A_1...{\d A_k\over\d\phi_l}
...(\theta A_m)...A_n\Bigr)-$$
$$-i\delta (y-x_k)T\Bigl( A_1...{\d (\theta A_k)\over\d\phi_l}...A_n\Bigr)
\Bigr\}
\Delta_{lj}(x_k-z)+i\d_\mu^y T\Bigl({\d j^\mu\over\d\phi_l}(y)A_1...A_n\Bigr)
\Delta_{lj}(y-z).\eqno(B.9)$$

For $\phi_j\not=\psi,\psq$ the last term vanishes and we obtain zero, due to
$${\d(\theta A_k)\over\d\phi_l}\Delta_{lj}=
\theta\Bigl({\d A_k\over\d\phi_l}\Bigr)\Delta_{lj}\quad\quad({\rm for}\quad
\phi_j\not=\psi,\psq)\eqno(B.10)$$ 
and the inductive assumption.

If $\phi_j=\psq$ ($\phi_j=\psi$ is analogous) the last term is equal to
$$i\d_\mu^y\Bigl\{ T\Bigl(\psq (y)A_1...A_n\Bigr)\gamma^\mu S(y-z)\Bigr\}
=-i\sum_k T\Bigl( A_1...{\d A_k\over\d\psi}...A_n\Bigr)\delta (x_k-y)S(x_k-z)
\eqno(B.11)$$
according to {\bf (N4)}. Because of
$${\d(\theta A_k)\over\d\psi}=i(r_k-s_k){\d A_k\over\d\psi},\quad\quad
\theta\Bigl({\d A_k\over\d\psi}\Bigr)=i(r_k-1-s_k){\d A_k\over\d\psi}
\eqno(B.12)$$
and the inductive assumption, the commutator (B.9) vanishes also in this case.

Again we conclude that a possible violation of a Ward identity 
(we call it an anomaly)
can only appear in the vacuum sector, i.e. in the vacuum expectation values. 
$$a(y,x_1,...,x_n)\=d\d_\mu^y T\Bigl( j^\mu (y)A_1(x_1)...A_n(x_n)\Bigr)-$$
$$-i\sum_{j=1}^n\delta (y-x_j)
T\Bigl( A_1(x_1)...(\theta A_j)(x_j)...A_n(x_n)\Bigr)=$$
$$=\d_\mu^y <\Omega|T\Bigl( j^\mu (y)A_1...A_n\Bigr)|\Omega>-
i\sum_{j=1}^n\delta (y-x_j)
<\Omega|T\Bigl( A_1...(\theta A_j)...A_n\Bigr)|\Omega>.\eqno(B.13)$$
Moreover the anomalies are local, i.e.
$$a(y,x_1,...,x_n)=P(\d)\delta (x_1-y)...\delta (x_n-y),\eqno(B.14)$$
where $P(\d)$ is a polynomial in $\d\equiv (\d_{x_1},...,\d_{x_n})$.
The latter is a consequence of the induction with respect to the order $n$ and
the causal factorization (2.2) of the time ordered products.

{\it Step 2:} Next we prove the Ward identities with a free field factor. 
We only need
to consider their vacuum expectation values. The normalization 
condition {\bf (N4)}
implies the well known identity
$$<\Omega|T(A_1(x_1)...A_n(x_n)\phi_i(x))|\Omega>=$$
$$=i\sum_{k=1}^n\sum_l\Delta^F_{il}(x-x_k)
<\Omega|T(A_1(x_1)...{\d A_k\over\d\phi_l}(x_k)...A_n(x_n))|\Omega>,
\eqno(B.15)$$
where $\Delta^F$ is the Feynman propagator.
By inserting this formula we obtain
$$\d_\mu^y <\Omega|T\Bigl( j^\mu (y)A_1(x_1)...A_m(x_m)\phi_i(x)\Bigr)|\Omega>
-i\sum_{j=1}^n\delta (y-x_j)
<\Omega|T\Bigl( A_1(x_1)...(\theta A_j)(x_j)...$$
$$...A_m(x_m)\phi_i(x)\Bigr)|\Omega>
-i\delta (y-x)<\Omega|T\Bigl( A_1(x_1)...A_m(x_m)(\theta \phi_i)(x)\Bigr)|
\Omega>=$$
$$=i\sum_k\Delta^F_{il}(x-x_k)\d_\mu^y <\Omega|T\Bigl( j^\mu (y)A_1...
{\d A_k\over\d\phi_l}...A_m\Bigr)|\Omega>+$$
$$+i\d_\mu^y\Bigl\{\Delta^F_{il}(x-y)
<\Omega|T\Bigl({\d j^\mu\over\d\phi_l}(y)A_1...A_m\Bigr)|\Omega>\Bigr\}+$$
$$+\sum_j\delta (y-x_j)\sum_{k\,(k\not= j)}\Delta^F_{il}(x-x_k)
<\Omega|T\Bigl( A_1...
{\d A_k\over\d\phi_l}...(\theta A_j)...A_m\Bigr)|\Omega>+$$
$$+\sum_k\delta (y-x_k)\Delta^F_{il}(x-x_k)<\Omega|T\Bigl( A_1...
{\d (\theta A_k)\over\d\phi_l}...A_m\Bigr)|\Omega>+$$
$$+\delta (y-x)\sum_k\Delta^F_{(\theta i)l}(x-x_k)<\Omega|T\Bigl( A_1...
{\d A_k\over\d\phi_l}...A_m\Bigr)|\Omega>,\eqno(B.16)$$
where $m=n-1$ and $\Delta^F_{(\theta i)l}(x-y)\=d i<\Omega|T((\theta\phi_i)(x)
\phi_l(y))|\Omega>$.

If $\phi_i\not=\psi,\psq$ the second and the last term vanish 
($\alpha_i=0$). Due
to (B.10) and the Ward identities in lower order we get zero.

If $\phi_i=\psq$ ($\phi_i=\psi$ is analogous) the second term is equal to
$$i\sum_k<\Omega|T\Bigl( A_1...{\d A_k\over\d\psi}...A_m\Bigr)|\Omega>
[S^F(x_k-y)\delta (y-x)-\delta (x_k-y)S^F(y-x)]\eqno(B.17)$$
by means of {\bf (N4)}. Because of (B.12) and the Ward identities in 
lower order all
terms cancel in this case, too.

{\it Step 3:} By choosing $g$ as in (B.4) we conclude from Lemma 8
$$0=\int d^4y\,g(y)a(y,x_1,...,x_n)=\int d^4y\,a(y,x_1,...,x_n)\eqno(B.18)$$
in ${\cal D}'({\bf R}^{4n})$. This restricts the remaining anomalies.
We want to remove them by finite renormalizations
of $<\Omega|T\Bigl( j^\mu (y)A_1(x_1)...A_n(x_n)\Bigr)|\Omega>$. 
This can only be done
if the polynomials $P(\d)$ (B.14) have the form
$$P(\d)=(\sum_{i=1}^n\d_\mu^{x_i})P_1^\mu (\d),\quad\quad P_1^\mu (\d)\quad
{\rm polynomial}\quad {\rm in}\quad  \d\equiv (\d_{x_1},...,\d_{x_n}).
\eqno(B.19)$$
To prove this we consider the Fourier transformation of the anomaly (B.14),
$$\hat a(y,p_1,...,p_n)=(2\pi)^{-2n}\int dx_1...dx_n\,a(y,x_1,...,x_n)
e^{i(p_1x_1+...+p_nx_n)}=$$
$$=(2\pi)^{-2n}P(-ip_1,....,-ip_n)e^{i(p_1+...+p_n)y}.\eqno(B.20)$$
From (B.18) we know that $P(-ip_1,....,-ip_n)$ vanishes on the submanifold
$\sum_{i=1}^np_i=0$,
$$P(-ip_1,....,-ip_n)\delta(\sum_{i=1}^np_i)=0.\eqno(B.21)$$
Let $\tilde P(q,p_1,...,p_{n-1})\=d P(-ip_1,....,-ip_n)$, 
where $q\=d \sum_{i=1}^np_i$.
We consider the Taylor series of $\tilde P$
$$\tilde P(q,p_1,...,p_{n-1})=\sum_{k=1}^{{\rm degree}\>\tilde P}\sum_{|\alpha|
+|\beta|=k}{q^\alpha p^\beta\over \alpha !\beta !}
\Bigl({\d^{|\alpha|}\over\d q^\alpha}
{\d^{|\beta|}\over\d p^\beta}\tilde P\Bigr)(0),\quad p\equiv (p_1,...,p_{n-1}).
\eqno(B.22)$$
The terms $|\alpha|=0$ vanish because $\Bigl({\d^{|\beta|}\over\d p^\beta}
\tilde P\Bigr)(0)$ is obtained by varying $\tilde P$ on the submanifold
$q=0$. There remain only terms with a factor $q^\alpha,\>|\alpha|\geq 1$.
This proves (B.19).

{\it Step 4:} But there is still a problem. The renormalization
$$<\Omega|T\Bigl( j^\mu (y)A_1(x_1)...A_n(x_n)\Bigr)|\Omega>\rightarrow
<\Omega|T\Bigl( j^\mu A_1...A_n\Bigr)|\Omega>+P_1^\mu (\d)\delta (x_1-y)...
\delta (x_n-y),\eqno(B.23)$$
(which removes the anomaly) is only admissible if 
$P_1^\mu (\d)\delta (x_1-y)...
\delta (x_n-y)$ has the same symmetries as required for
$<\Omega|T\Bigl( j^\mu A_1...A_n\Bigr)|\Omega>$. 
Especially if there are factors $j^{\mu_l}(x_l)$ among $A_1(x_1)...A_n(x_n)$
the permutation symmetry with respect to $(y,\mu)\leftrightarrow (x_l,\mu_l)$
must be maintained (for all $l$). There is a prominent counterexample 
where this
is impossible: the axial anomaly, i.e. $<\Omega|T\Bigl( j_A^\mu (y)
j_A^{\mu_1}(x_1)
j_A^{\mu_2}(x_2)\Bigr)|\Omega>$, where $j_A^\mu\=d 
:\psq\gamma^\mu\gamma^5\psi:$.

We have not found a general argument (for non-axial QED) that all 
possible anomalies
factorize $P(\d)=(\sum_i\d_\mu^{x_i})P_1^\mu (\d)$ such that $P_1^\mu (\d)$ has
the wanted symmetries. However, taking step 2 into account,
and also the fact that the scaling degree at $x_j-y=0\>(\forall j=1,...,n)$
[BF] of the anomaly
cannot be higher than the scaling degree of the terms in the corresponding Ward
identity, the number of Ward identities which can be violated is 
strongly reduced.
In addition, due to (B.18) terms of singular order $\omega =0$ 
(i.e. scaling degree
$\delta =4n$) are excluded in the anomalies. The famous Ward identity 
which connects
the vertex function with the electron self-energy has only one factor $j$ and,
hence, the renormalization (B.23) maintains the symmetries in 
that case. There remain
the following anomalies
$$\d_\mu^y <\Omega|T\Bigl( j^\mu (y){\cal L}(x_{11})...{\cal L}(x_{1m})
j^{\mu_1}(x_{21})\Bigr)|\Omega>=
\sum_{1\leq |a|\leq 3}C_{1a}^{\mu_1}\d^a\prod_{h,j}
\delta (x_{hj}-y),\eqno(B.24)$$
$$\d_\mu^y <\Omega|T\Bigl( j^\mu (y){\cal L}(x_{11})...{\cal L}(x_{1m})
j^{\mu_1}(x_{21})j^{\mu_2}(x_{22})j^{\mu_3}(x_{23})\Bigr)|\Omega>=$$
$$=\sum_{|a|= 1}C_{2a}^{\mu_1\mu_2\mu_3}\d^a\prod_{h,j}\delta (x_{hj}-y).
\eqno(B.25)$$
The unknown constants $C_{l...}^{...},\>l=1,2,3,4$ are restricted by Lorentz
covariance and the permutation symmetry in $x_{11},...,x_{1m}$. 
The analogous Ward identity with three factors $j$ is trivially 
fulfilled, due to
Furry's theorem, by imposing C-invariance as a further normalization condition.
\footnote{$^{11}$}{In the inductive construction of the time ordered products 
C-invariance
can only get lost in the extension to the total diagonal, because of the causal
factorization (2.2). Starting with an extension which fulfills all 
other normalization
condition {\bf (N1-4)}, {\bf (N6)}, (B.6) and symmetrizing it with respect to 
C-invariance, we obtain an extension which satisfies all requirements.}
The anomalies in (B.24) and (B.25) can be further restricted
by the symmetry in the factors $j$ of the terms on the l.h.s., e.g.
$\d_{\mu_1}^{x_{21}}\d_\mu^y<\Omega|T\Bigl(j^\mu (y){\cal L}(x_{11})...
{\cal L}(x_{1m})
j^{\mu_1}(x_{21})\Bigr)|\Omega>$ is symmetrical in $y,x_{21}$. 
By working this out
one finds that the factorization (B.19) of the anomalies can be done 
in such a way that
the symmetries are preserved in the renormalizations 
(B.23) [DHS2]. $\quad\quad\w$
\vskip 1cm
{\bf Acknowledgements}: We profitted from discussions with Franz-Marc Boas, 
Izumi Ojima, Klaus Sibold and Raymond Stora which are gratefully acknowledged.
Part of this work was done at the Max-Planck-institute
for mathematics in the sciences in Leipzig and at the university of Leipzig.
The authors wish to thank Bodo Geyer, Gert Rudolph, Klaus Sibold and 
Eberhard Zeidler for warm hospitality.
\vfill
\eject
{\trm References}
\vskip 1cm
[BF] R.Brunetti and K.Fredenhagen:"Interacting quantum fields in curved 
space: Re\-normalization of $\phi^4$", gr-qc/9701048, {\it Proceedings of 
the Conference 'Operator Algeras 
and Quantum Field Theory', held at Accademia Nazionale dei Lincei, Rome, 
July 1996}

R.Brunetti and K.Fredenhagen:"Microlocal analysis and interacting 
quantum field 
theories: Renormalization on physical backgrounds", in preparation

[BDF] F.-M.Boas, M.D\"utsch and K.Fredenhagen:"A local (perturbative) 
construction of observables in gauge theories: 
nonabelian gauge theories", work in progress

[BFK] R.Brunetti, K.Fredenhagen and M.K\"ohler:"The microlocal spectrum 
condition and Wick polynomials of free fields on curved spacetimes", 
{\it Commun. Math. Phys.} {\bf 180} (1996), 312

[BRS] C.Becchi, A.Rouet and R.Stora:"Renormalization of the abelian 
Higgs-Kibble model", {\it Commun. Math. Phys.} {\bf 42} (1975), 127

C.Becchi, A.Rouet and R.Stora:"Renormalization of gauge theories", 
{\it Annals of Physics (N.Y.)} {\bf 98} (1976), 287

[BlSe] P.Blanchard and R.Seneor:"Green's functions for theories with 
massless particles (in perturbation theory)", {\it Ann. Inst. H. 
Poincar\'e A} {\bf 23} (1975), 147

[BS] N.N.Bogoliubov and D.V.Shirkov, "Introduction to the Theory of Quantized 
Fields", New York (1959)

[Bu1] D.Buchholz, M.Porrmann and U.Stein:"Dirac versus Wigner: 
Towards a universal 
particle concept in local quantum field theory", {\it Phys. Lett. B} 
{\bf 267} (1991), 377

[Bu2] D.Buchholz:"Gauss' law and the infraparticle problem", 
{\it Phys. Lett. B} {\bf 174} (1986), 331

[BW] M.Bordemann and S.Waldmann:"Formal GNS construction and states in 
deformation quantization", q-alg/9611004, 
to appear in {\it Commun. Math. Phys.}

[D1] M.D\"utsch:"On gauge invariance of Yang-Mills theories with 
matter fields", {\it N. Cimento A} {\bf 109} (1996), 1145

[D2] M.D\"utsch:"Slavnov-Taylor identities from the causal point of view", 
{\it Int. J. Mod. Phys. A} {\bf 12} (1997) 3205

[DF] M.D\"utsch and K.Fredenhagen:"Deformation stability of BRST-quantization",
hep-th/9807215, to appear in the proceedings of the conference 
'Particles, Fields and Gravitation', Lodz, Poland (1998)

[DHS1] M.D\"utsch, T.Hurth and G.Scharf:"Causal construction of 
Yang-Mills theories. IV. Unitarity", {\it N. Cimento A} {\bf 108} (1995), 737

[DHS2] M.D\"utsch, T.Hurth and G.Scharf:"Gauge invariance of massless QED", 
{\it Phys. Lett. B} {\bf 327} (1994), 166

[DHKS1] M.D\"utsch, T.Hurth, F.Krahe and G.Scharf:"Causal construction of 
Yang-Mills theories. I." {\it N. Cimento A} {\bf 106} (1993), 1029

[DHKS2] M.D\"utsch, T.Hurth, F.Krahe and G.Scharf:"Causal construction of 
Yang-Mills theories. II." {\it N. Cimento A} {\bf 107} (1994), 375

[DKS] M.D\"utsch, F.Krahe and G.Scharf:"Interacting fields in finite QED", 
{\it N. Cimento A} {\bf 103} (1990), 871

[DS] M.D\"utsch and G.Scharf:"Perturbative gauge invariance: the 
electroweak theory", hep-th/9612091

A.Aste, M.D\"utsch and G.Scharf:"Perturbative gauge invariance: the 
electroweak theory II", hep-th/9702053

[EG] H.Epstein and V.Glaser:"The role of locality in perturbation theory", 
{\it Ann. Inst. H. Poincar\'e A} {\bf 19} (1973), 211

[F] R.P.Feynman, {\it Acta Phys. Polonica} {\bf 24} (1963), 697

[FHRW] J.S.Feldman, T.R.Hurd, L.Rosen and J.D.Wright:"QED: A Proof of
Renormalizability", Springer-Verlag (1988)

[FP] L.D.Faddeev and V.N.Popov:"Feynman diagrams for the Yang-Mills field", 
{\it Phys. Lett. B} {\bf 25} (1967), 29

[K] F.Krahe:"A causal approach to massive Yang-Mills theories", 
{\it Acta Phys. Polonica B} {\bf 27} (1996), 2453

[KO] T.Kugo and I.Ojima:"Local covariant operator formalism of 
non-abelian gauge 
theories and quark confinement problem", {\it Suppl. Progr. 
Theor. Phys.} {\bf 66} (1979), 1

[R] M.Requardt:"Symmetry conservation and integrals over local 
charge desities in 
quantum field theory", {\it Commun. Math. Phys.} {\bf 50} (1976), 259

[S] G.Scharf:"Finite Quantum Electrodynamics", 2nd. ed., Springer-Verlag (1995)

[Sch] B.Schroer:"Infrateilchen in der Quantenfeldtheorie", 
{\it Fortschr. Phys.} {\bf 173} (1963), 1527

[St1] R.Stora:"Local gauge groups in quantum field theory: perturbative gauge 
theories", talk given at the workshop 'Local Quantum Physics' at the 
Erwin-Schroedinger-institute, Vienna (1997)

[St2] R.Stora:"Lagrangian field theory", summer school of theoretical physics 
about 'particle physics', Les Houches (1971)

[St3] R.Stora:"Differential algebras in Lagrangean field theory", 
ETH-Z\"urich Lectures, January-February 1993; 

G.Popineau and R.Stora:"A pedagogical remark on the main theorem of 
perturbative renormalization theory", unpublished preprint

[Ste] O. Steinmann,"Perturbation expansions in axiomatic field theory", 
Lecture Notes in Physics {\bf 11}, Springer-Verlag (1971)

[V] R. Verch,"Local Definiteness, Primarity and Quasiequivalence of 
Quasifree Hada\-mard Quantum States in Curved Spacetime", 
{\it Commun. Math. Phys.} {\bf 160} (1994), 507

\bye